\documentclass[12pt,showpacs,twocolumn, preprintnumbers,aps,pre, reprint,superscriptaddress]{revtex4-1}

\usepackage{graphicx}
\usepackage{subfigure}
\usepackage{color}
\usepackage{amsmath}
\usepackage[linktocpage,colorlinks=true,linkcolor=blue,citecolor=blue,breaklinks=true]{hyperref}
\usepackage{verbatim}
\usepackage{breakcites}

\begin{document}

\preprint{}

\title{Fluctuations in Arctic Sea Ice Extent: Comparing Observations and Climate Models}

\author{Sahil Agarwal}
\affiliation{Yale University, New Haven, CT, USA}
\email[]{sahil.agarwal@yale.edu}

\author{John S. Wettlaufer}
\affiliation{Yale University, New Haven, CT, USA}
\affiliation{Mathematical Institute, University of Oxford, Oxford, UK}
\affiliation{Nordita, Royal Institute of Technology and Stockholm University, SE-10691 Stockholm, Sweden}
\email[]{john.wettlaufer@yale.edu}

\date{\today}

\begin{abstract}

The fluctuation statistics of the observed sea ice extent during the satellite era are compared with model output from CMIP5 models using a multi-fractal time series method.  The two robust features of the observations are that on annual to bi-annual time scales the ice extent exhibits white noise structure and there is a decadal scale trend associated with the decay of the ice cover.  It is shown that (i) there is a large inter-model variability in the time scales extracted from the models, (ii) none of the models exhibit the decadal time scales found in the satellite observations, (iii) {5} of the 21 models examined exhibit the observed white noise structure, and (iv) the multi-model ensemble mean exhibits neither the observed white noise structure nor the observed decadal trend.  It is proposed that the observed fluctuation statistics produced by this method serve as an appropriate test bed for modeling studies. 

\end{abstract}

\pacs{}

\maketitle

\section{Introduction} 

Polar amplification and the ice-albedo feedback focus scientific study on the fluctuations in the areal coverage of high latitude ice.  By area, { on average} the southern hemisphere ice cover is dominated by the Antarctic ice sheet, whereas in the northern hemisphere the Arctic sea ice cover dominates.  Although we can observe the daily ice cover from space, 
the substantial changes in Arctic ice mass during recent decades are associated with an ostensibly unmeasurable ($\sim$ 1 W m$^{-2}$) contribution to the surface energy balance \cite{OneWatt}.   Hence, given this sensitivity, developing a quantitative understanding of the fluctuations in sea ice cover is important.  According to the {Inter-governmental Panel on Climate Change (IPCC)} Fifth Assessment Report \cite{IPCC:2013}, the most reliably measured characteristic of sea ice is the hemispheric sea ice extent, to which models are tuned and parameterizations refined.  To this end we compare the statistical structure of satellite observations of ice extent to that from model output.  

The  Global Climate Models (GCMs) from the IPCC Assessment Report 5 (AR5) project the Arctic to be ice free by the middle of this century, whereas the previous assessment projected this to occur at the end of the century. 
The AR5 models project the Arctic to be ice-free as early as 2030 to as late as 2100 (\cite{Stroeve:2012aa}, Fig. 1 in \citet{Wang:2012}). The high intermodel variability and the  parameterization schemes and tuning used (e.g. the sea ice albedo, clouds, convective processes) \cite{EUW:2007, DeWeaver:2008, EUW:2008, Mauritsen:2012aa, Hourdin:2013aa} constitute key aspects of their veracity at projecting the state of the ice cover.  

To quantify the fluctuations in the ice cover from days to decades we have analyzed satellite passive microwave data using a multifractal methodology \cite{Sahil:MF}.  
We studied  the Arctic Equivalent Sea Ice Extent (EIE), where the EIE is defined as the total surface area, including land, north of the zonal-mean ice edge latitude, and thus is proportional to the sine of the ice edge latitude \cite{IanGeom}.  (By studying the EIE one minimizes coastal effects.) 
We find that the EIE is a multi-fractal in time \cite{Sahil:MF} and thus cannot be explained as an auto-regressive process with a single time scale (a so-called AR-1 process), as is commonly used to characterize Arctic sea ice in GCMs \cite{Piwowar:1996, Blanchard:2010aa, Armour:2011}.  An AR-1 process is inappropriate because ($a$) The existence of multiple time scales in the data cannot be treated in a quantitatively consistent manner with a single decay autocorrelation. ($b$) The strength of the seasonal cycle is such that, if not appropriately removed, retrievals will always produce a single characteristic time  of about a year; a time scale at which all moments converge. Hence, the upper bound on the persistence time in any study that assumes an AR-1 process will inevitably be $\sim$ 1 year, as is indeed found for ice area \cite{Blanchard:2010aa, Armour:2011}. Separate studies with GCMs \cite{Lindsay:2008, Wang:2012aa, Chevallier:2013aa,  Sigmond:2013aa, Tietsche:2014aa} also show that an upper bound exists on the predictability for the Arctic sea ice cover and that predictability is \emph{skillful} only on the seasonal time-scales. Therefore, our approach highlights the dangers of not carefully detrending the seasonality.

\begin{figure}[h!]
	\centering
	    \includegraphics[trim = 50 20 30 20, clip, width = 0.5\textwidth]{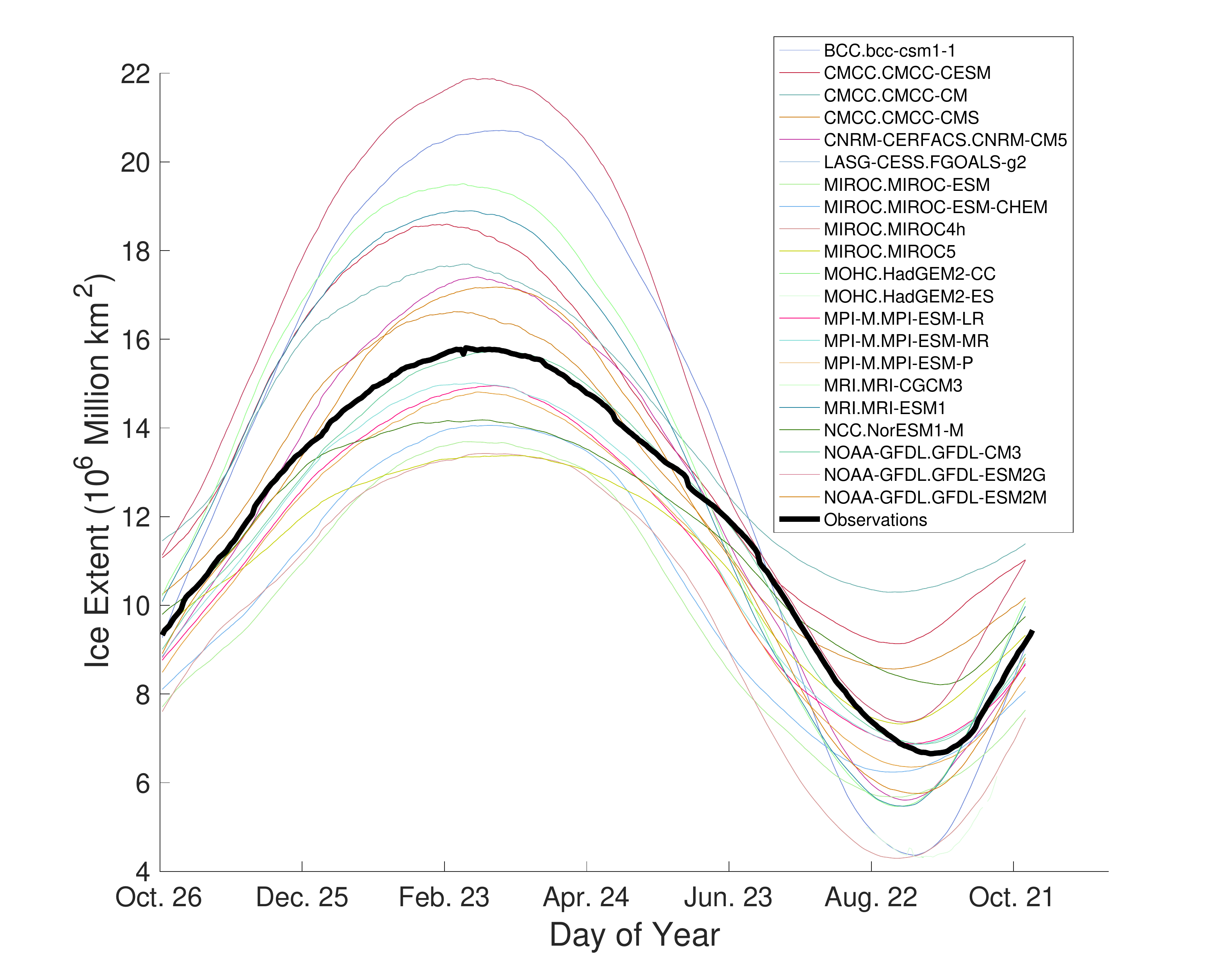}	  
      \caption{The mean seasonal cycle of the Ice Extent from the models in CMIP5 compared with the observations (in bold).}%
      \label{fig:Mean_IE}
\end{figure}

\begin{figure}[t]
		\centering
	   \includegraphics[trim = 50 20 30 10, clip, width = 0.5\textwidth]{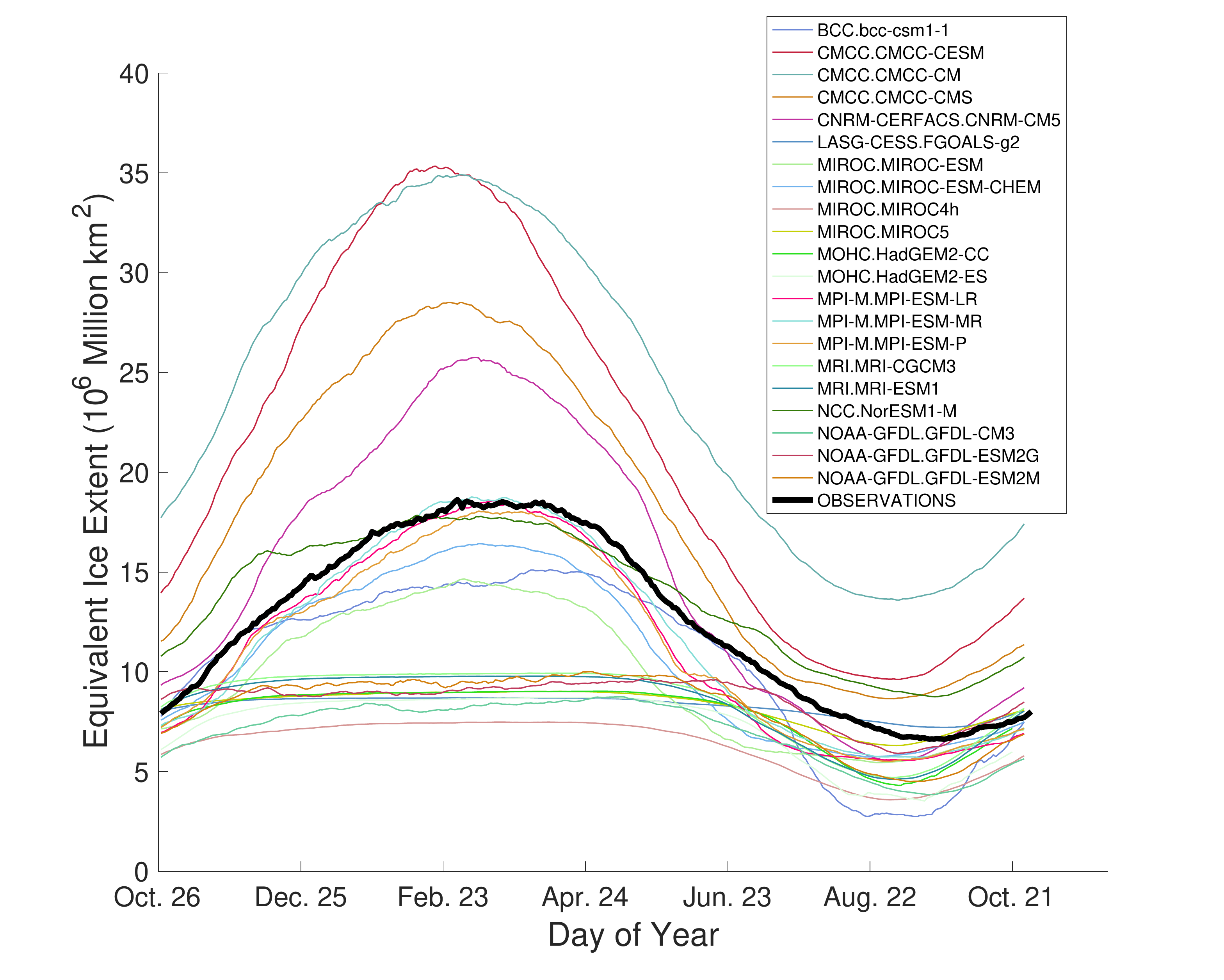}	  
      \caption{The mean seasonal cycle of the Equivalent Ice Extent from the models in CMIP5 compared with the observations (in bold).}%
      \label{fig:Mean_EIE}
\end{figure}

\begin{table*}
\centering
\caption{Global Climate Models (GCMs) used in this study. Models with daily output for sea-ice extent in their historical run have been included. { (* denotes models with two ensemble members included)}.}
\label{Tb:models}
\begin{tabular}{l l}
\hline
 {\bf Institution}  & {\bf Model Name}   \\
\hline 
Beijing Climate Center (BCC)                                            & BCC\_CSM1.1 (BCC Climate System \\
											&	Model, version 1.1) \\
Centro Euro-Mediterraneo per I Cambiamenti Climatici    & CMCC-CESM\\
Centro Euro-Mediterraneo per I Cambiamenti Climatici    & CMCC-CM \\
Centro Euro-Mediterraneo per I Cambiamenti Climatici    & CMCC-CMS \\
Centre National de Recherches Meteorologiques             & CNRM-CM5 (Coupled Global Climate\\
											&	 Model, version 5)*\\
Sate Key Laboratory of Numerical Modeling for 	      & FGOALS-g2 (Flexible Global Ocean-\\
Atmospheric Sciences and Geophysical Fluid Dynamics,		      &Atmosphere-Land System Model\\
Institute of Atmospheric Physics  					& gridpoint, version 2)\\
Atmosphere and Ocean Research Institute, UTokyo		& MIROC-ESM*\\
Atmosphere and Ocean Research Institute, UTokyo 		& MIROC-ESM-CHEM \\
Atmosphere and Ocean Research Institute, UTokyo  		& MIROC 4h* \\
Atmosphere and Ocean Research Institute, UTokyo 		& MIROC version 5 (MIROC5)*\\
Met Office Hadley Centre                                                    & HadGEM2-CC (Hadley Global \\
											& Environment Model 2 - Carbon Cycle)*\\
Met Office Hadley Centre                                                    & HadGEM2-ES (Hadley Global  \\
											& Environment Model 2 - Earth System)\\
Max Planck Institute for Meteorology (MPI-M)                     & MPI-ESM-LR (MPI Earth System Model, \\&Low Resolution)*\\
Max Planck Institute for Meteorology (MPI-M)                     & MPI-ESM-MR (MPI Earth System Model, \\&Medium Resolution)*\\
Max Planck Institute for Meteorology (MPI-M)                     & MPI-ESM-P (MPI Earth System Model, Paleo)*\\
Meteorological Research Institute (MRI)                              & MRI-CGCM3 (MRI Coupled Atmosphere-Ocean \\ &General Circulation Model, version 3)\\
Meteorological Research Institute (MRI)                              & MRI-ESM1 (MRI Earth System Model, version 1)\\
Norwegian Climate Centre                                                   & NorESM1-M (Norwegian Earth System Model,\\&  version 1, Medium resolution)*\\ 
Geophysical Fluid Dynamics Laboratory 		                  & GFDL CM3 (GFDL Climate Model, version 3)*\\   
Geophysical Fluid Dynamics Laboratory 		                  & GFDL ESM2G (GFDL Earth System Model)\\   
Geophysical Fluid Dynamics Laboratory  		                  & GFDL ESM2M (GFDL Earth System Model)\\
\hline 
\end{tabular}
\end{table*}

\section{Methods and Data}
In an effort to improve understanding of Arctic sea ice behavior in the AR5 GCMs, we compare the daily satellite observations from 1978-2005 to the model output data with daily frequency for the same period. For parity in this comparison, we analyze the 21 AR5 models (shown in Table \ref{Tb:models}) with daily data in their historical runs, and { use the first ensemble member from 11 of these and two from the remaining 10 models}.  To be consistent in comparison of all the models and the satellite observations, a common latitude-longitude grid of $0.5^{\circ} \times 0.5^{\circ}$ is used and we interpolate the sea ice concentration (or sea ice area fraction) from the original model/observation grid onto this interpolated grid. As is done for the observations, the sea ice extent obtained from the models is then converted to EIE.

Next, we remove the mean seasonal cycle (the mean EIE on each calendar day of the year) from the model output. Figures \ref{fig:Mean_IE} and \ref{fig:Mean_EIE} show the mean seasonal cycle for the Arctic sea ice extent and the EIE for the AR5 models along with the observations. The inter-model variability is remarkable. It is noted in the IPCC report \emph{Evaluation of Climate Models} \cite{IPCC:2013} that models are tuned to match the observed climate system, which in turn helps to give skillful predictions.  However, as is evident in Fig. \ref{fig:Mean_EIE}, where although the models may be tuned to match the Arctic sea ice extent (Fig. \ref{fig:Mean_IE}), if one calculates the EIE from the model output, the variation from the observed seasonal cycle is substantial, not only in the magnitude of the mean seasonal cycle, but also in the qualitative shape of the curves. Namely, some models have a \emph{cycloidal structure}, with a plateau shaped region in winter and a deep well structure in summer. We emphasize again that the EIE not only mitigates the effects of land on the ice extent, but it also appropriately characterizes regional sea-ice characteristics. Thus, even if a model is able to faithfully reproduce the total ice extent, it is possible that regional differences between the models and the observations are large and summing these acts to minimize these differences \cite{Li:2017aa}, thereby leading to specious conclusions. 

When viewing a time series as a temporal multifractal, there is no constraint on the number of allowable timescales present within the data.  Thus, depending on the temporal resolution of the time series, one can extract all  timescales corresponding to the physical processes governing the system.   \citet{Kantelhardt:2002}  developed a multi-fractal generalization of detrended fluctuation analysis (a modification of the rescaled range method used by \citet{Hurst:1951aa} to study the dynamics of river discharge) called multi-fractal detrended fluctuation analysis (MF-DFA). We use a new variant of MF-DFA called multi-fractal temporally weighted detrended fluctuation analysis (MF-TW-DFA), which exploits the intuition that points closer in time are more likely to be related than points distant in time, providing a clearer signature of long timescales present in a time series \cite{Zhou:2010}. The detailed algorithm for MF-TW-DFA can be found in (\cite{Sahil:MF, Sahil:SIV, Sahil:EXO} and refs. therein) and this is the approach we use to analyze and compare the EIE (without a seasonal cycle) from AR5 models to that from the satellite observations. Because of the fact that there are no \emph{a-priori} assumptions made regarding the physical processes in the system, this method has also been used to explain the dynamics of Arctic sea ice velocity fields \cite{Sahil:SIV}, the detection of exoplanets \cite{Sahil:EXO} and their atmospheres \cite{Sahil:ExoAtmos}. Note that when the seasonal cycle is not removed from the data multifractality is inhibited and hence all the time scales longer than annual are masked, which leads to erroneous interpretation of data or model output as obeying an AR-1 process. 

The fluctuations in a time series are quantified with respect to a smooth {\em profile}, defined as the cumulative sum produced by time-weighting the data. These fluctuations are then combined to construct the fluctuation function, $F_q(s)$, for each timescale, $s$, under examination, where $q$ denotes the $q^{{th}}$ moment of this function.  For a given moment, the key behavior examined is the $s$-dependence of $F_q (s)$, which is characterized by a generalized Hurst exponent $h(q)$ viz.,
\begin{equation}
F_q (s) \propto s^{h(q)} .  
\label{eq:power}
\end{equation}
For example, in a monofractal timeseries $h(q)$ is independent of $q$, and thus equivalent to the classical Hurst exponent $H$. The exponent $h(2)$ is related to the decay of the power spectrum, $S(f)$.  Hence, if $S(f) \propto f^{- \beta}$, with frequency $f$, then $h(2) = (1 + \beta)/2$ (e.g., \cite{Ding}). Thus, for white noise $\beta = 0$, which gives $h(2) = 1/2$, whereas for red noise $\beta = 2$, giving $h(2) = 3/2$.  Therefore, the dominant timescales found using MF-TW-DFA  
also capture the corresponding temporal dynamics, such as white noise, red noise, and correlation structure, which allows us to construct stochastic models for the observed processes, such as the statistical structure and dynamics of sea ice velocity fields \cite{Sahil:SIV}. The ``crossover'' or ``dominant'' time is defined as that at which the fluctuation function $\log_{10} F_2(s)$ changes slope with respect to $\log_{10} s$.    These occur when slope of the curve exceeds a threshold of $C_{th} = 0.01$.

\section{Discussion}

\begin{figure}
		\centering
	    \includegraphics[width = 0.5\textwidth]{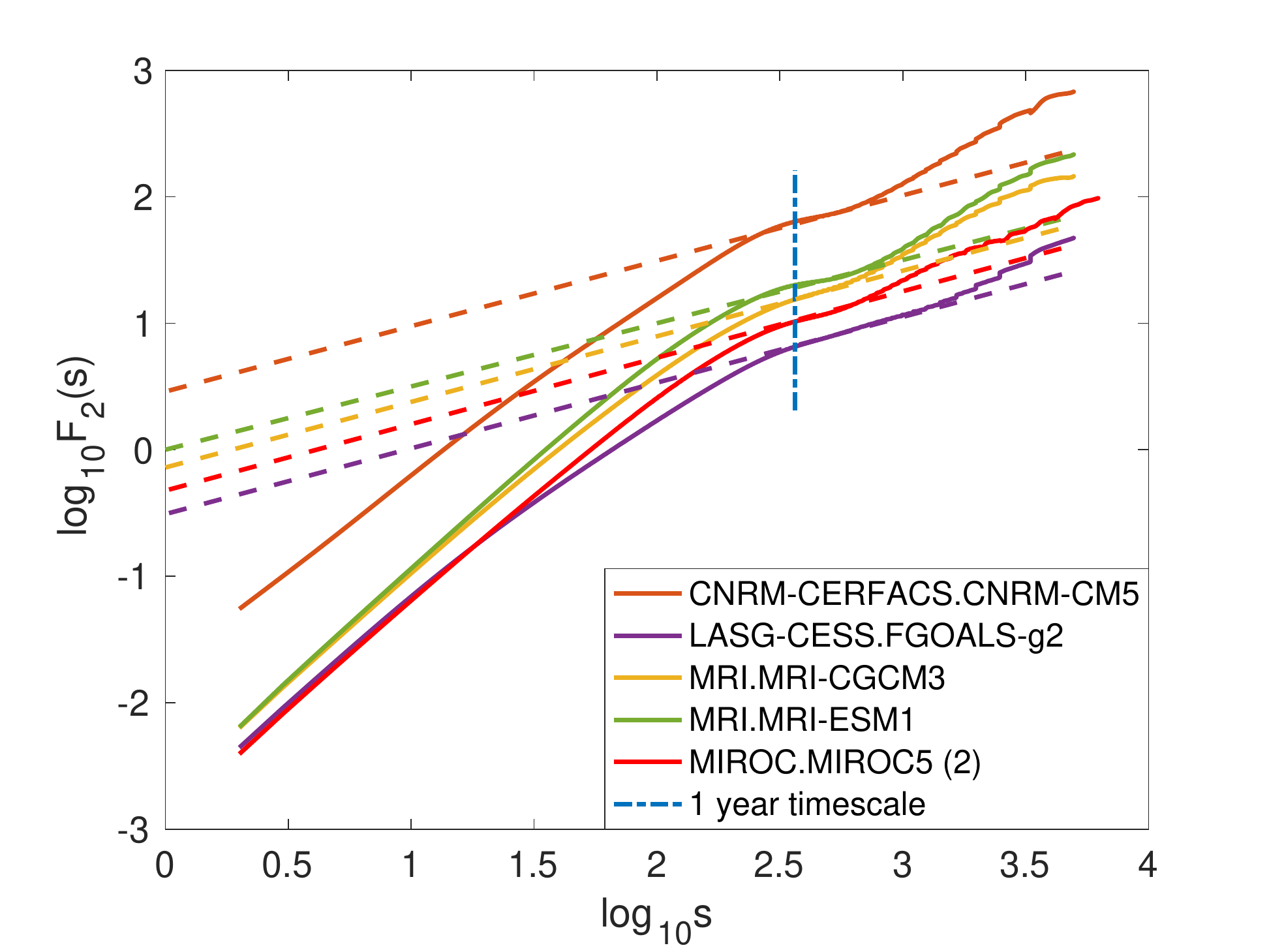}	  
      \caption{For $q$ = 2, the fluctuation function for EIE for the {five (of the 21 examined)} GCMs labeled in the inset. The slanted straight dashed lines denote white noise with $h$(2)=1/2, the vertical line denotes 1 year.  The time range is 1 day $\le s \le$ 13.6 years.  }%
      \label{fig:WhiteNoise_GCM}
     \vspace{-0.55 cm}
\end{figure}

Here we compare the fluctuation functions from the AR5 models to the observations. The nature and number of figures are such that the majority of the results are presented in the Supplementary Materials section, and here we summarize the key features, which are (1) large inter-model variability in the time scales extracted from the models, (2) {none} of the models exhibit the decadal time scales found in the satellite observations, and (3) only { 5 (CNRM-CM5, MRI-CGCM3 , MRI-ESM1, FGOALS-g2, MIROC5 (2nd ensemble member))} of the 21 models exhibit the white noise structure (shown in Fig. \ref{fig:WhiteNoise_GCM}) observed in the satellite record on annual to bi-annual time scales. Moreover, only 4 models exhibit time scales approaching decadal (7.2 years in BCC1-1, 4.8 years in FGOALS-g2, 7.9 years in NorESM1, { 4.5 years in NorESM1 (2nd ensemble member)}, and 6 years in GFDL-ESM2G), whereas in all of the other models the longest time scale is of the order a year. 

The high frequency processes governing shorter time scales are associated with the influence of synoptic systems, the intermediate time scales are seasonal and the longer -- decadal -- time scales are generally ascribed to climate processes \cite{Sahil:MF}.  For example, \citet{Sahil:SIV} demonstrated that on annual-biannual timescales, the sea ice extent is largely controlled by the wind fields over the Arctic. By comparing the sea-ice volumes from the models with the PIOMAS \cite{Zhang:2003aa} dataset, \citet{Shu:2015aa} showed that the sea-ice thickness produced in the CMIP5 models is much less than what is observed.  Therefore, although analysis of the Arctic sea-ice velocity fields from the CMIP5 models shows that most models reproduce the statistical characteristics of the observed velocity fields, if the sea ice produced by the models is too thin, the spectrum of its response to wind forcing will change.  For example, \citet{Colony:1980} showed that local low-frequency ice motion is linearly correlated with local synoptic wind forcing, but the local inertial motion is not, despite a high coherency between the inertial motion of the ice on the same length ($\sim$ 100 km) scales.  This highlights the role of mechanical interactions between ice floes on short time scales, which would not be captured in models producing thinner ice.   

\citet{Wang:2012} analyzed the sea-ice extent in the CMIP5 models and concluded that the Arctic would be ice-free in the next 14--36 years with a median of 28 years, based on the model spread.  \citet{Massonnet:2012aa} tried to reduce the CMIP5 uncertainty for the Arctic to become ice free, finding a spread of about 40-60 years.  
However, the absence of the longer time scales in the models that we have shown here provides a challenge for their predictive capability on decadal time scales. The origin of this lack of parity between satellite observations and the models will of course differ from model to model, but should provide an observational and statistical constraint for model physics, which has evidently not been met by the tuning and parameterization schema used thus far.  
Indeed, \citet{Swart:2015aa} showed how a deliberately biased picking of short-term trends to predict the long-term behavior of sea-ice extent can lead to false conclusions. Similarly, our study demonstrates both quantitatively and qualitatively the reason behind such high inter-model spread on decadal timescales and longer.

\begin{figure}
		\centering
	    \includegraphics[width = 0.5\textwidth]{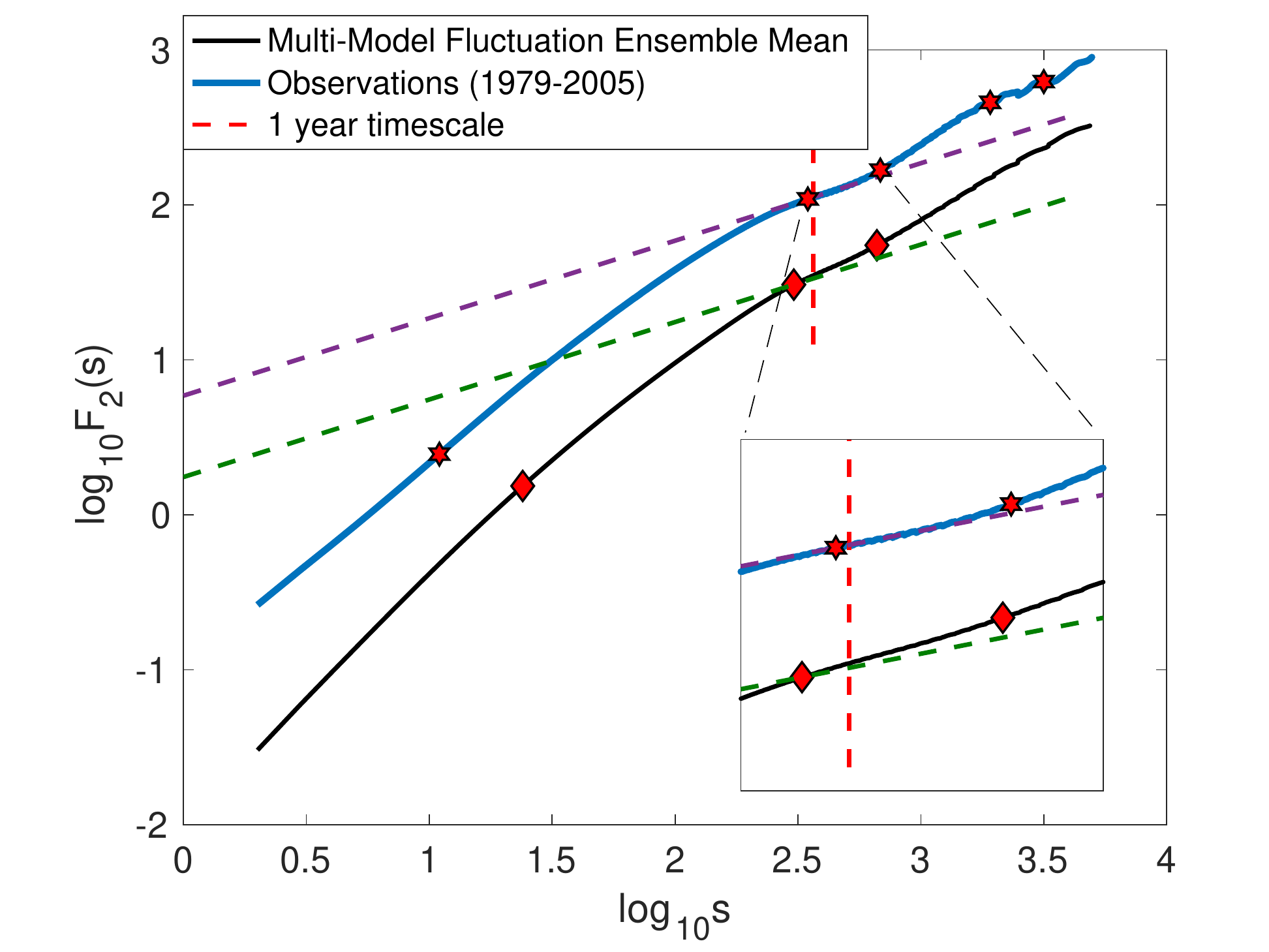}	  
      \caption{For $q$ = 2, the fluctuation functions for EIE from the Multi-Model Ensemble Mean (black) and the satellite observations (blue). The slanted straight dashed lines denote white noise with $h$(2)=1/2, the vertical line denotes 1 year. The time range is 1 day $\le s \le$ 13.6 years. Diamonds denote timescales for the MMEM (24 days, 304 days and 658 days), and hexagrams denote timescales for the satellite observations (11 days, 347 days, 684 days, 5.2 years and 8.6 years).}%
      \label{fig:MF_MMEM}
      \vspace{-0.65 cm}
\end{figure}

\section{Conclusion}

The observed Arctic sea ice extent can be thought of as the ``output'' of the highly complex nonlinear interactions governing the air/sea/ice system.  It is a central goal of modeling to try and reproduce this output.  
A model can be tuned to match some of the observations, but if one is interested in making predictions about the future state of the system, one seeks to know the key underlying physical and dynamical processes and how they govern the central statistical response of the system.  Many studies compare the multi-model ensemble mean (MMEM)
to observations, and argue that a sufficient condition for predictability is that the observations are within one standard deviation of the MMEM.  However, here we find that even the MMEM does not exhibit the robust observational features of white noise characteristics on annual-biannual timescales or the characteristic decadal timescales. Figure \ref{fig:MF_MMEM} compares the fluctuation functions for the EIE from the satellite observations (blue) to the MMEM. The longest timescale present in the MMEM data is 1.8 years and that from the observations is 8.6 years. Therefore, simply summing different models to produce the MMEM does not provide the observed statistical structure of the ice pack.  

A central role of observations is to produce robust statistical metrics that can serve as a target for modeling. Whereas, even with incorrect physics, it is possible to match features of the observations, capturing the appropriate statistical properties should provide the litmus tests for models.  Our method is agnostic with regard to the number of timescales, and hence associated processes, that may be present in the system. Therefore, we hope that the fact that we capture both the shortest and longest observed processes in the ice pack will provide a useful and robust test bed for modeling studies. 

\acknowledgements
The support of NASA Grant NNH13ZDA001N-CRYO is acknowledged by both authors. J.S.W. acknowledges Swedish Research Council grant no. 638-2013-9243 and a Royal Society Wolfson Research Merit Award. Data for the CMIP5 models is publicly available and the sea-ice satellite observations are available from NSIDC.

\appendix

\section{\label{sec:app}}

\begin{figure*}
\centering
   (a)\includegraphics[width = 0.45\textwidth]{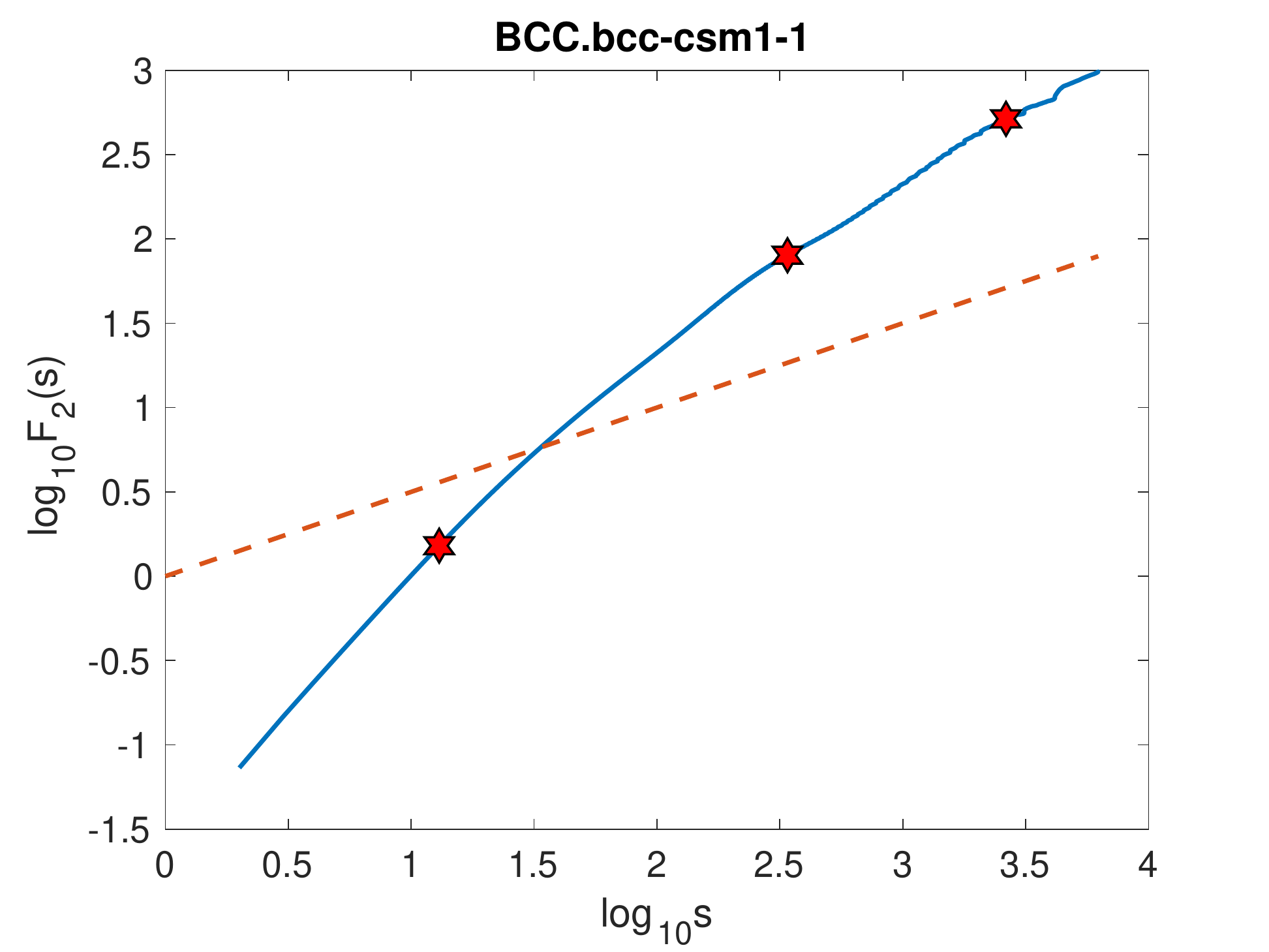}
   (b)\includegraphics[width = 0.45\textwidth]{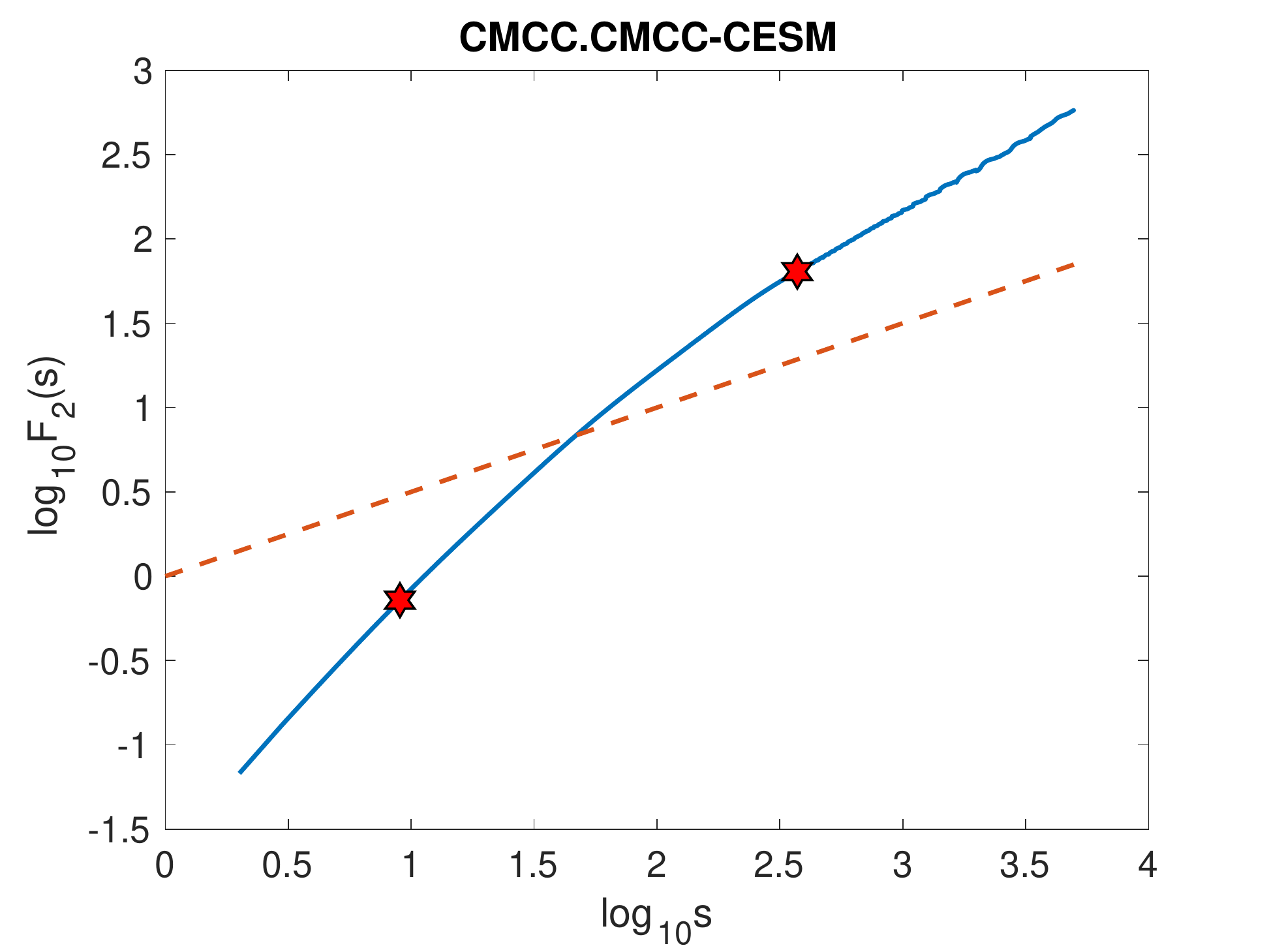}
   
   (c)\includegraphics[width = 0.45\textwidth]{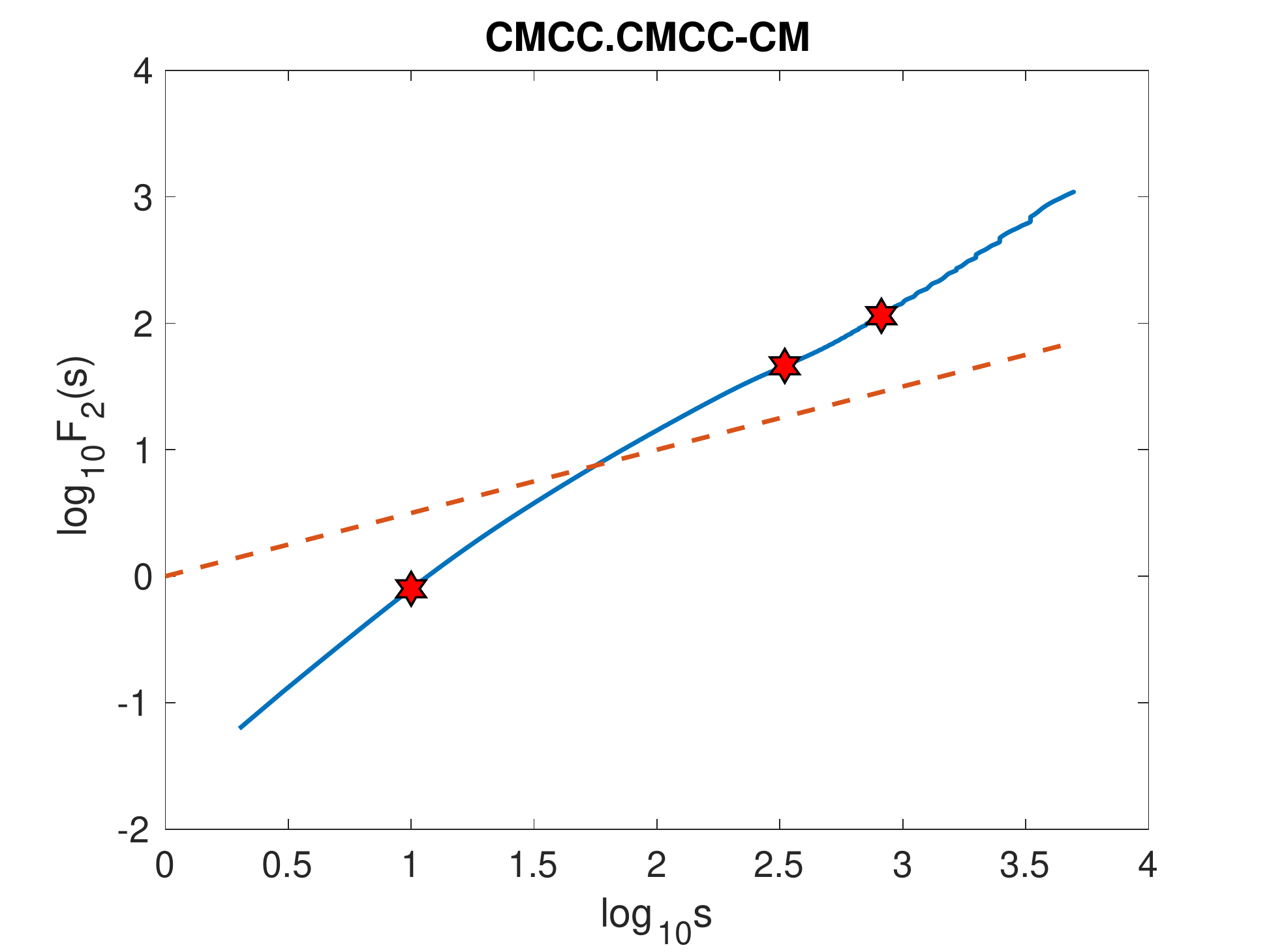}
   (d)\includegraphics[width = 0.45\textwidth]{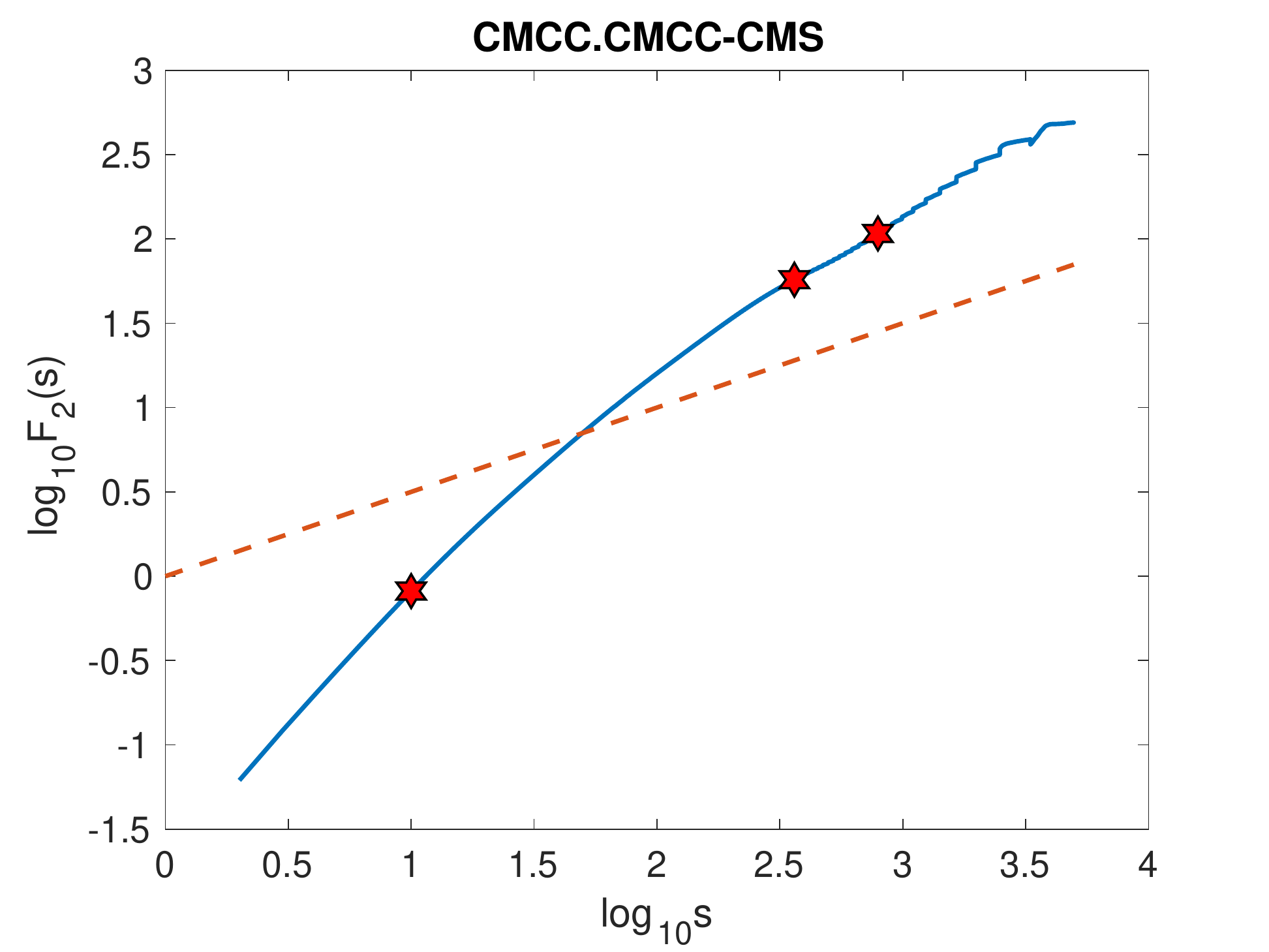}
   
   (e)\includegraphics[width = 0.45\textwidth]{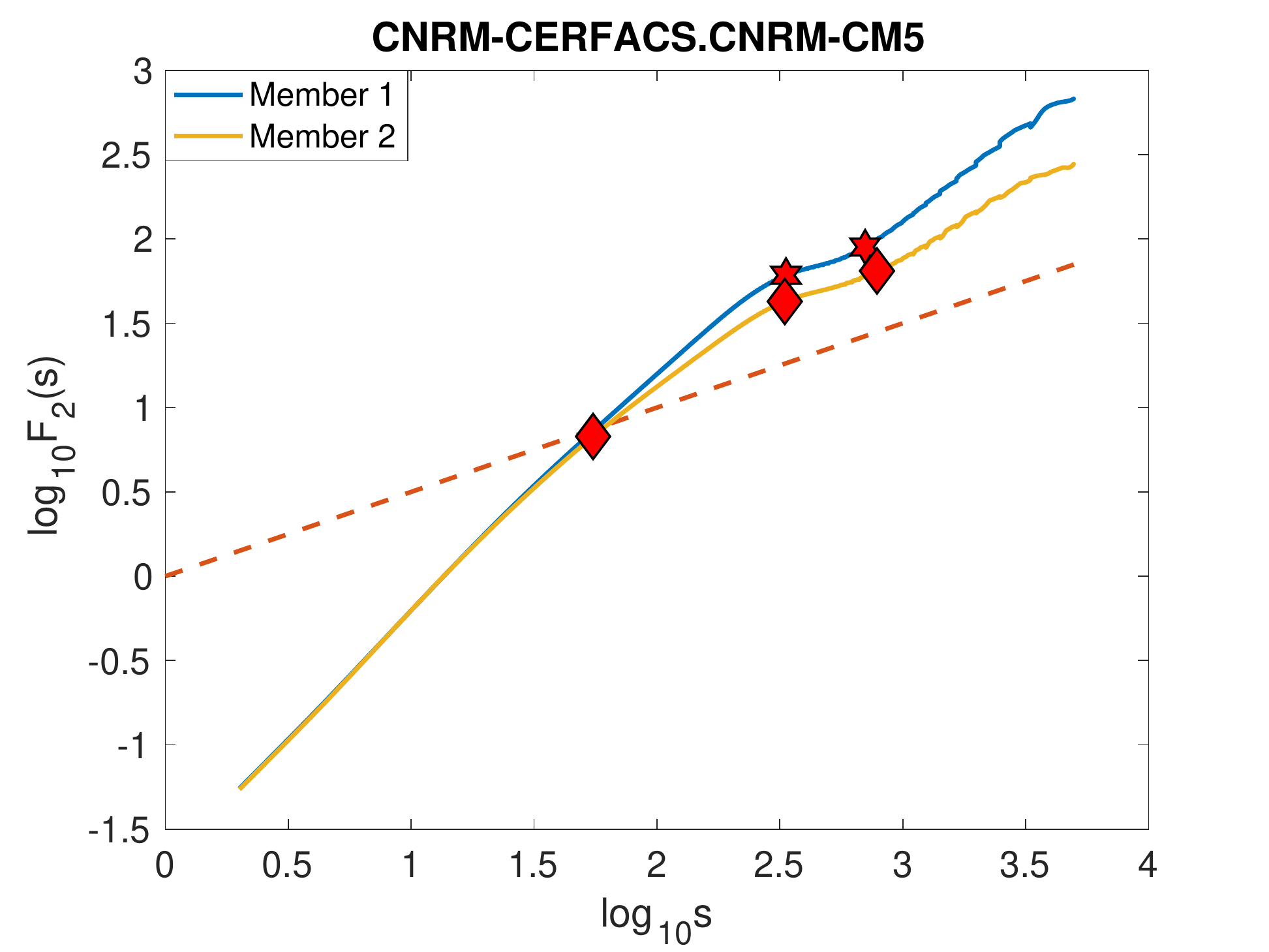}
   (f)\includegraphics[width = 0.45\textwidth]{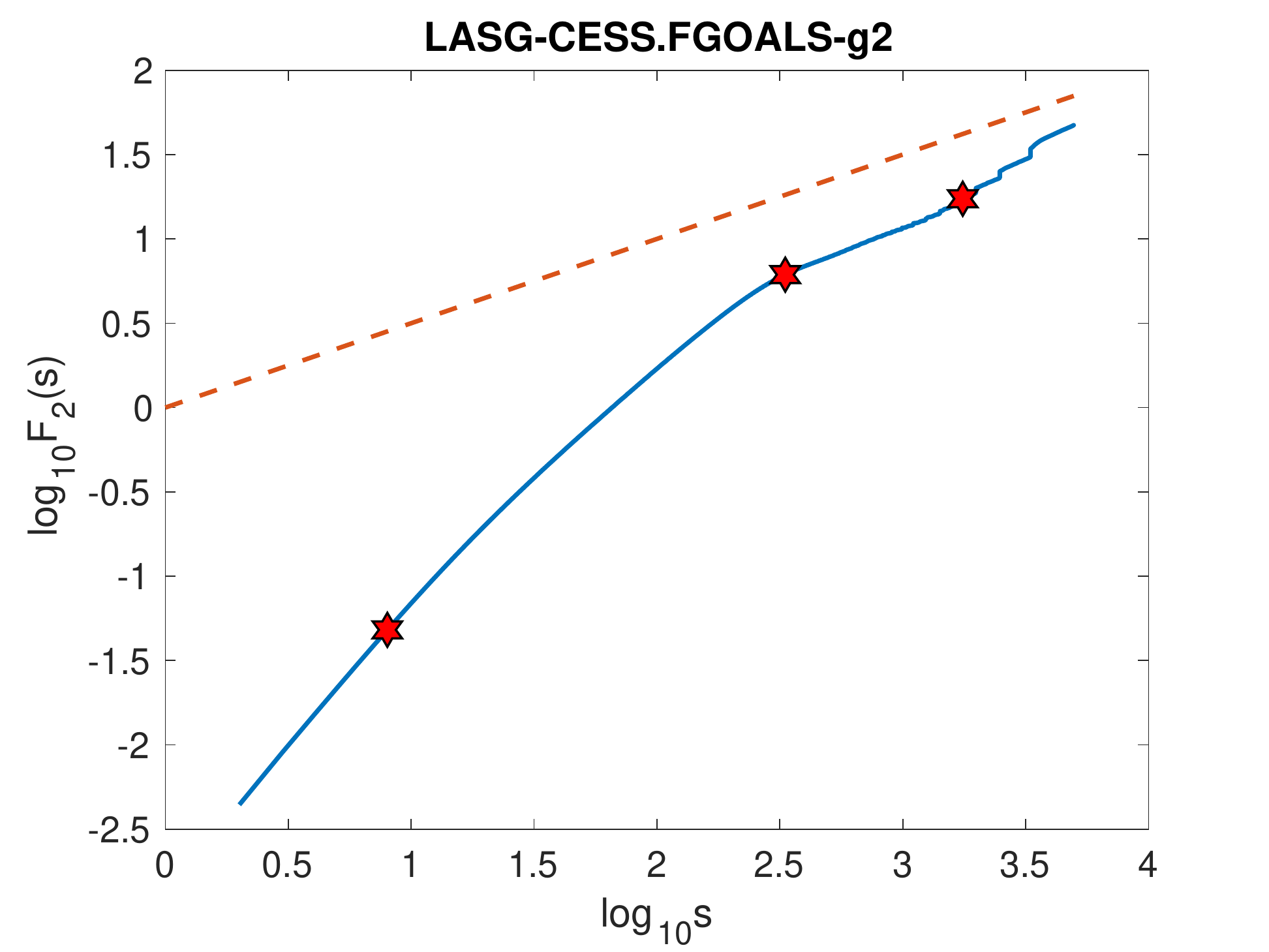}   
\caption{Fluctuation functions (blue) for CMIP5 AR5 models. The respective timescales present are (a) 13 days, 340 days, 7.2 years (stars); (b) 9 days, 372 days (stars); (c) 10 days, 331 days, 818 days (stars); (d) 10 days, 359 days, 793 days (stars); (e) { Member 1: 335 days, 703 days (stars); Member 2: 55 days, 331 days, 785 days (diamonds);} (f) 8 days, 333 days, 4.8 years (stars). The dashed line (orange) denotes a slope of 1/2.}
\label{fig:GCM_1}
\end{figure*}

\begin{figure*}
\centering
   (a)\includegraphics[width = 0.45\textwidth]{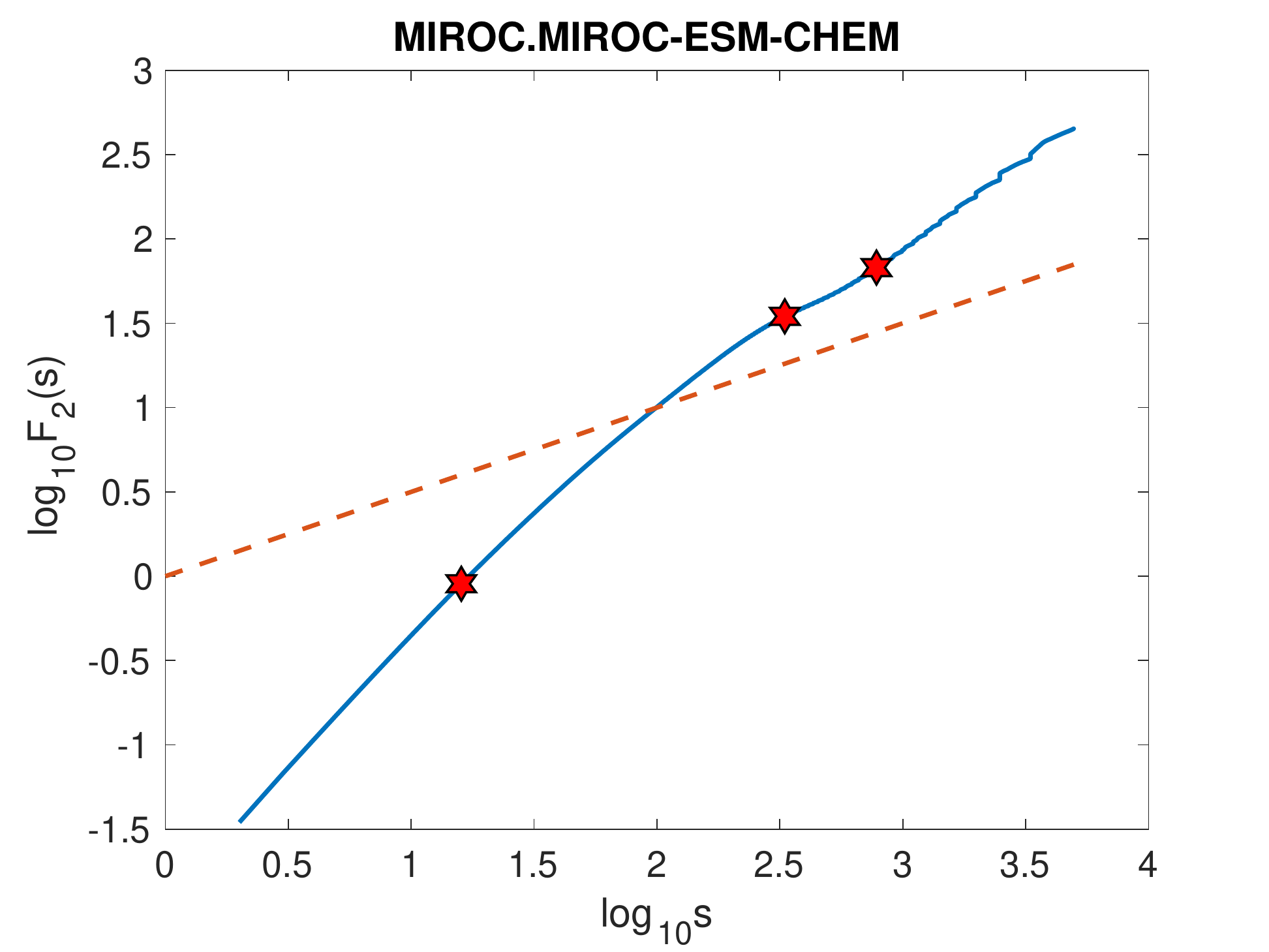}
   (b)\includegraphics[width = 0.45\textwidth]{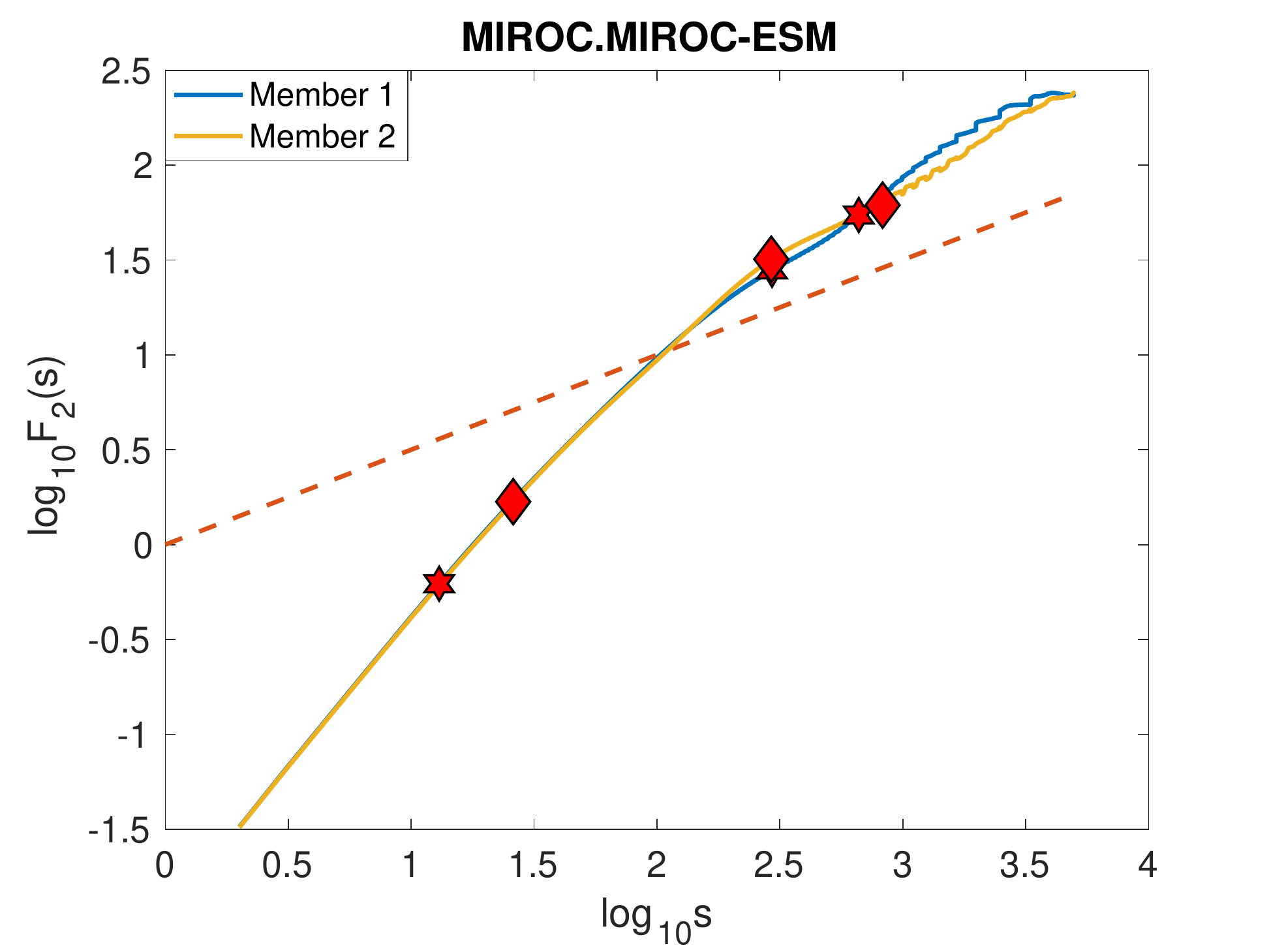}
   
   (c)\includegraphics[width = 0.45\textwidth]{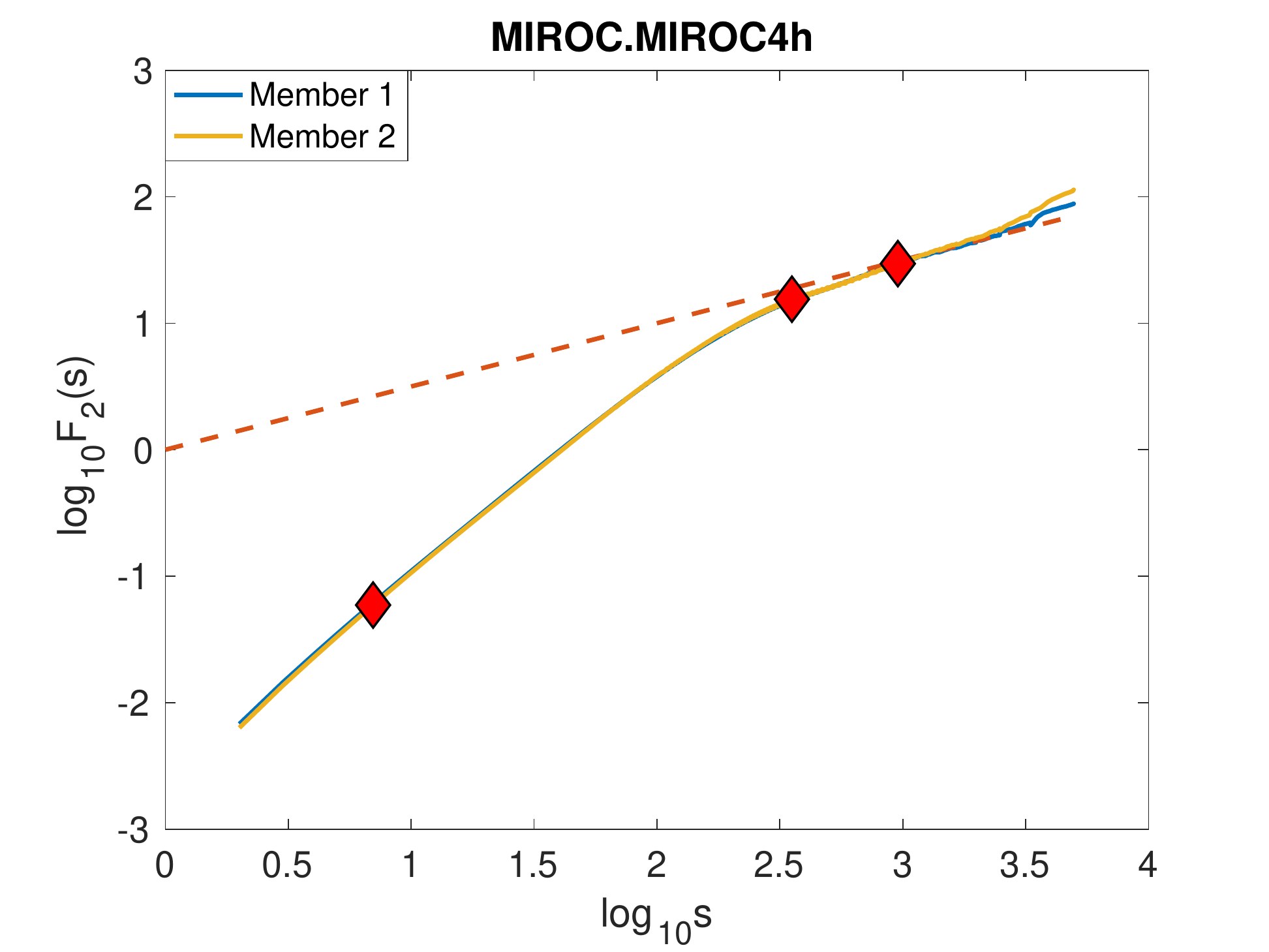}
   (d)\includegraphics[width = 0.45\textwidth]{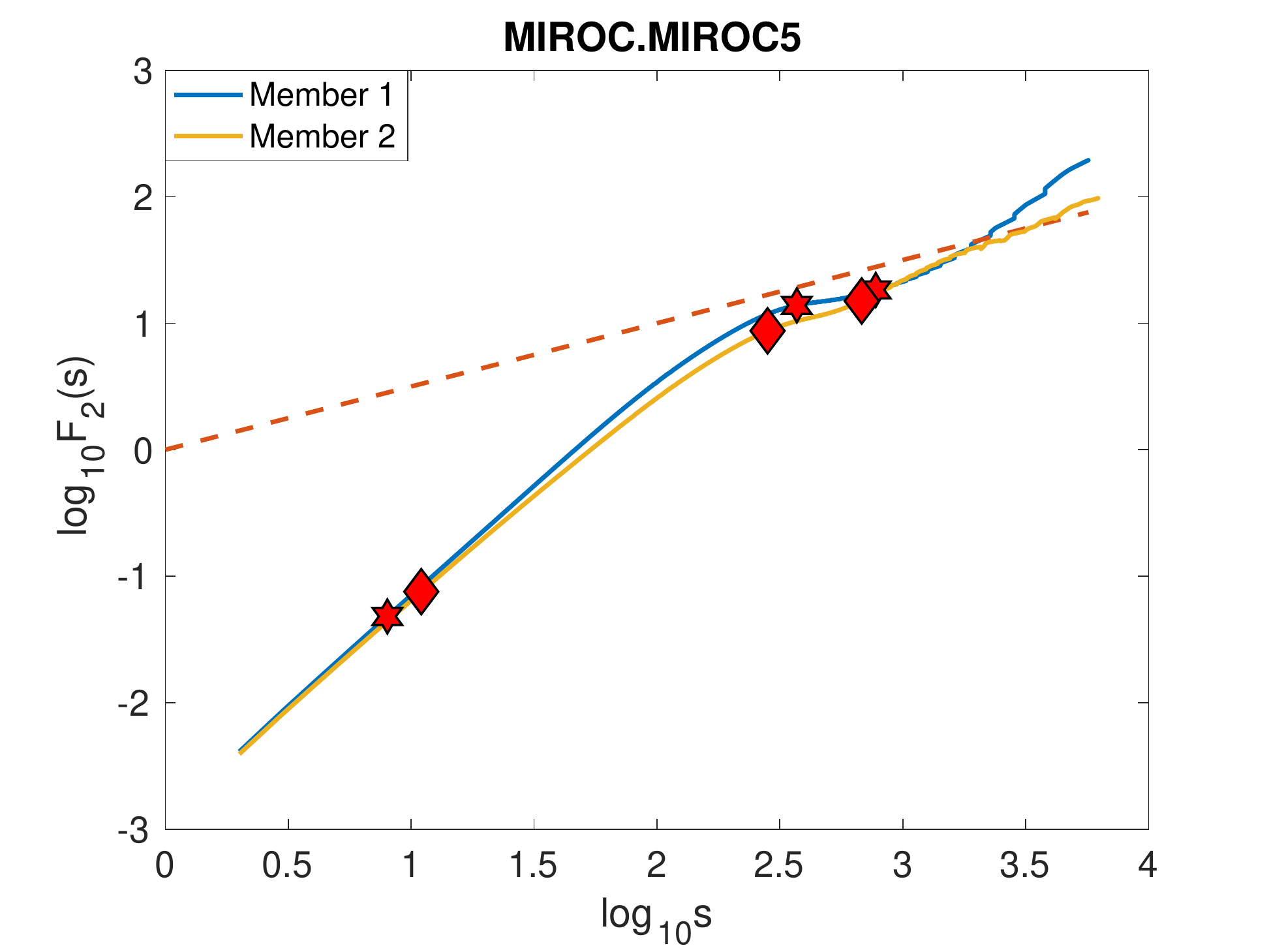}
   
   (e)\includegraphics[width = 0.45\textwidth]{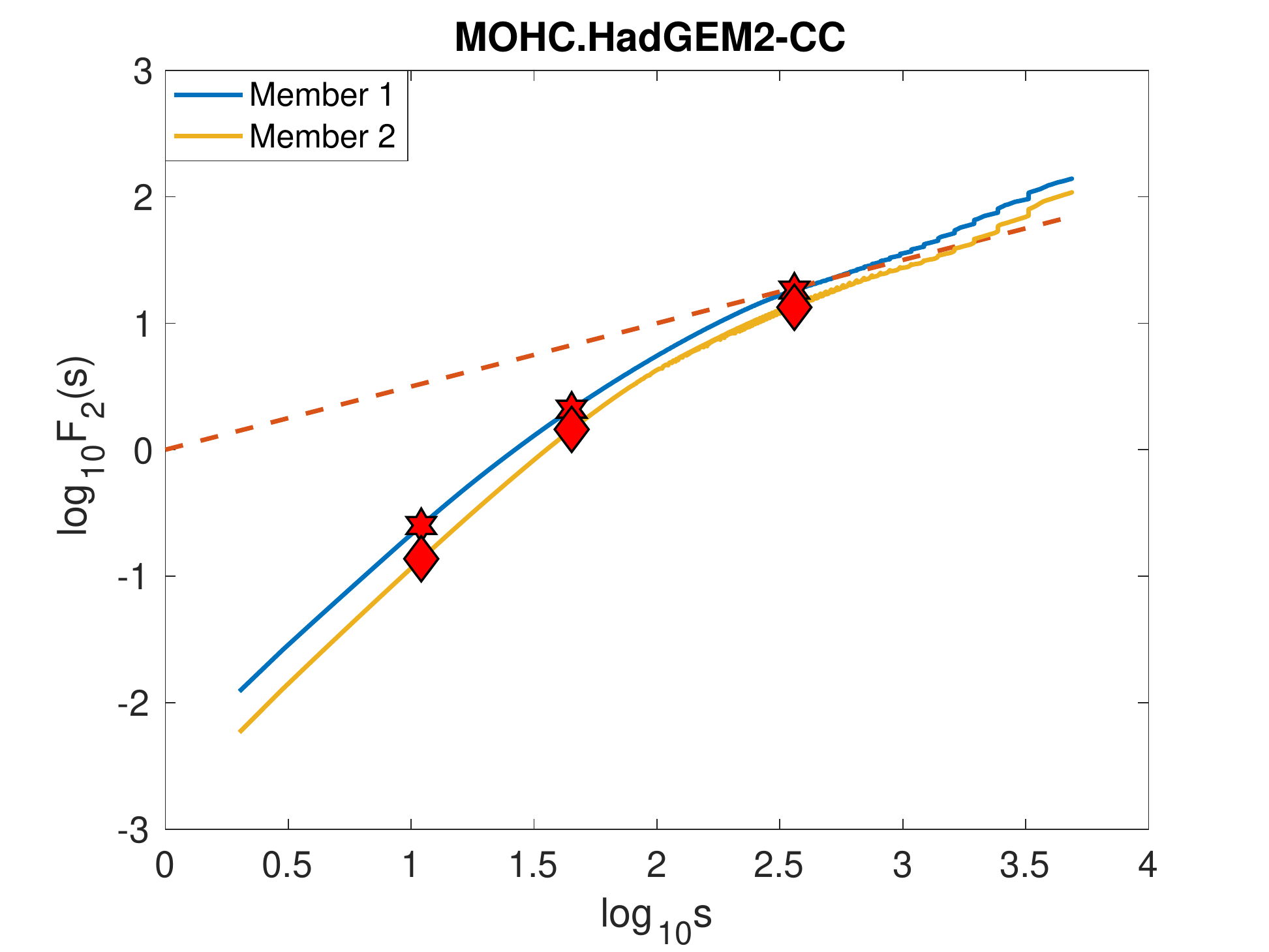}
   (f)\includegraphics[width = 0.45\textwidth]{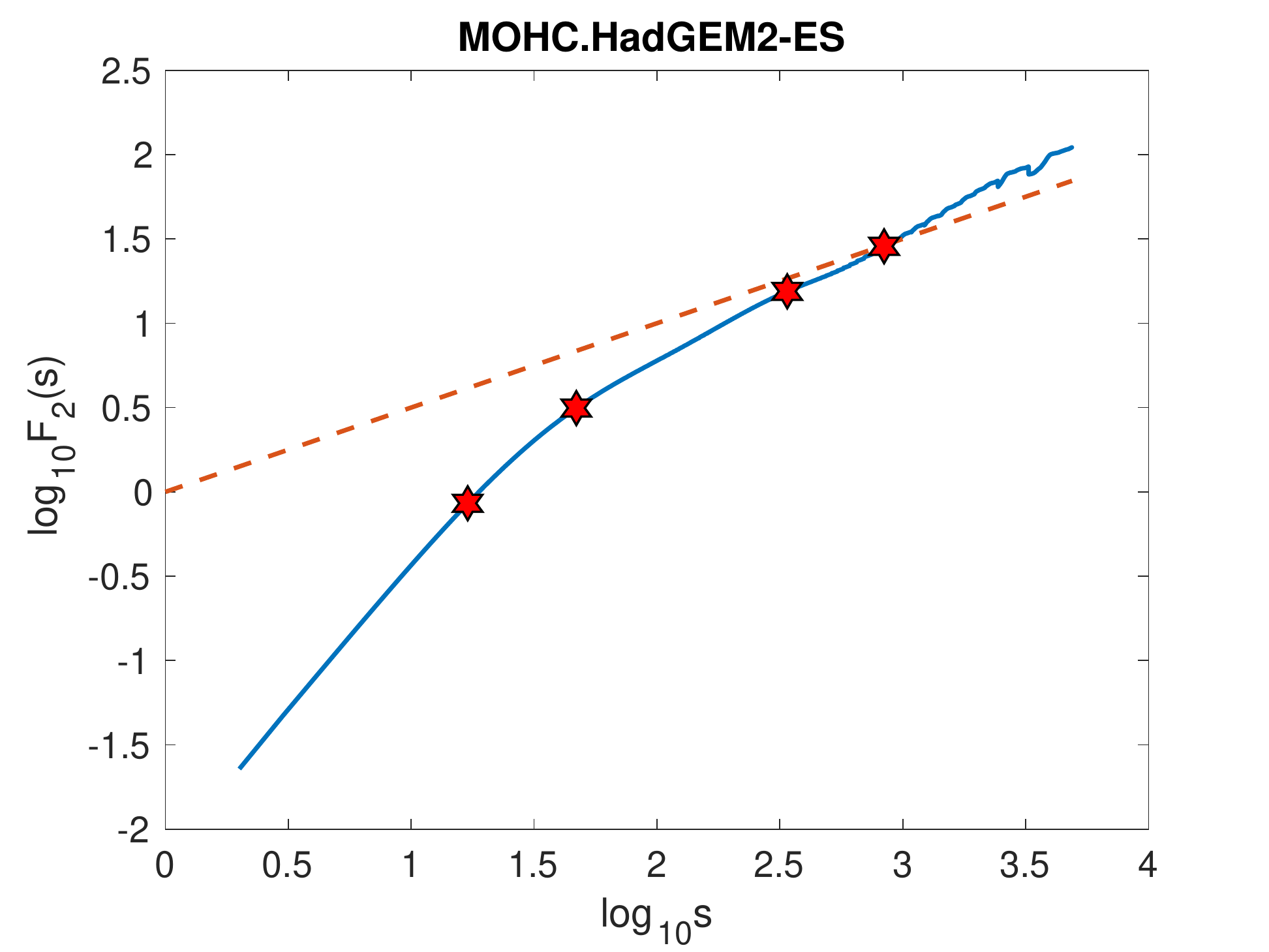}   
\caption{Fluctuation functions (blue) for CMIP5 AR5 models. The respective timescales present are (a) 16 days, 330 days, 782 days (stars); (b) { Member 1: 13 days, 294 days, 661 days (stars); Member 2: 26 days, 292 days, 828 days (diamonds); (c) Member 1: 7 days, 354 days, 955 days (stars); Member 2: 7 days, 354 days, 955 days (diamonds); (d) Member 1: 8 days, 371 days, 777 days (stars); Member 2: 11 days, 282 days, 681 days (diamonds); (e) Member 1: 11 days, 45 days, 362 days (stars); Member 2: 11 days, 45 days, 362 days (diamonds); }(f) 17 days, 47 days, 339 days, 840 days (stars). The dashed line (orange) denotes a slope of 1/2.}
\label{fig:GCM_2}
\end{figure*}

\begin{figure*}
\centering
   (a)\includegraphics[width = 0.45\textwidth]{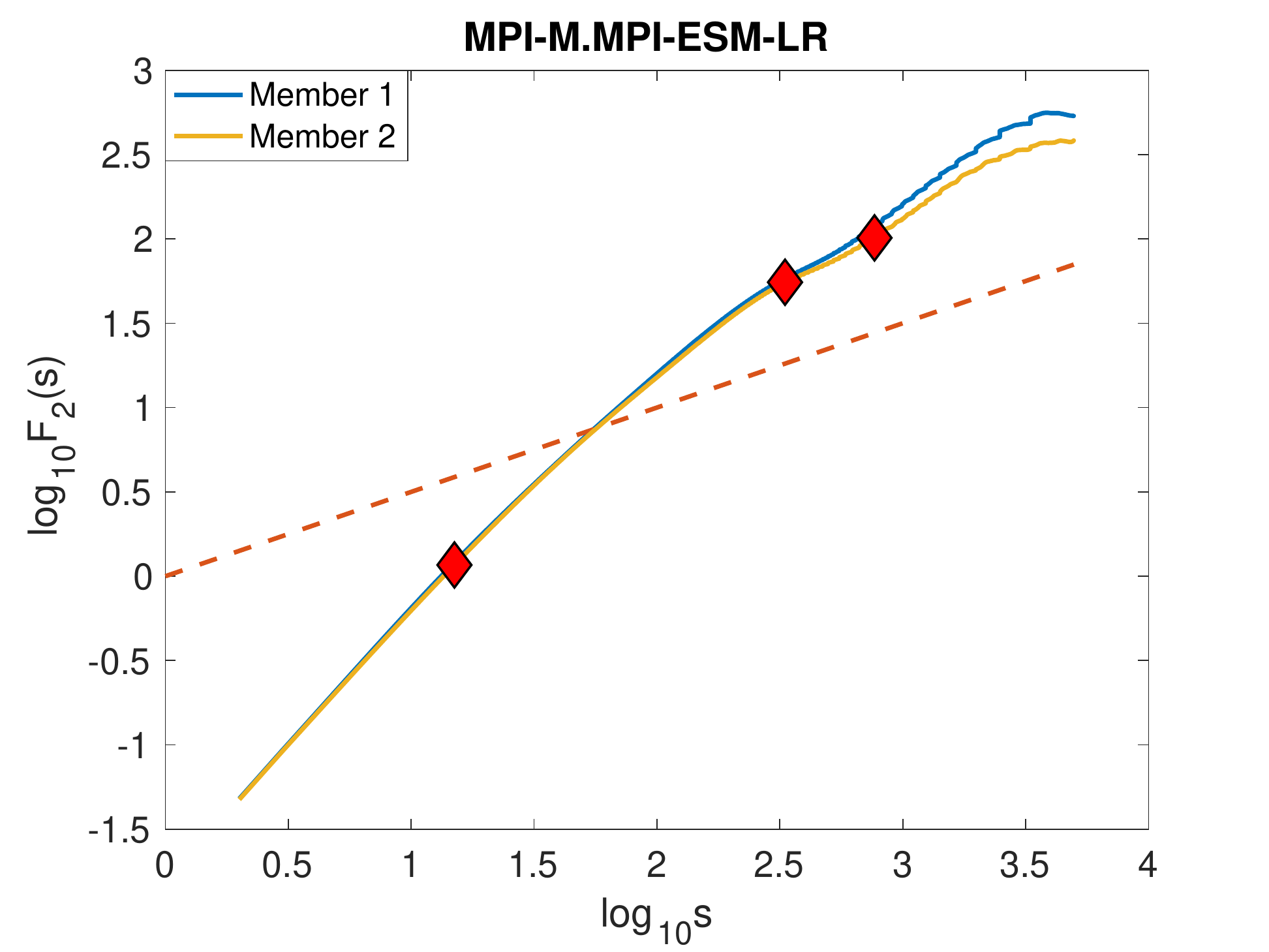}
   (b)\includegraphics[width = 0.45\textwidth]{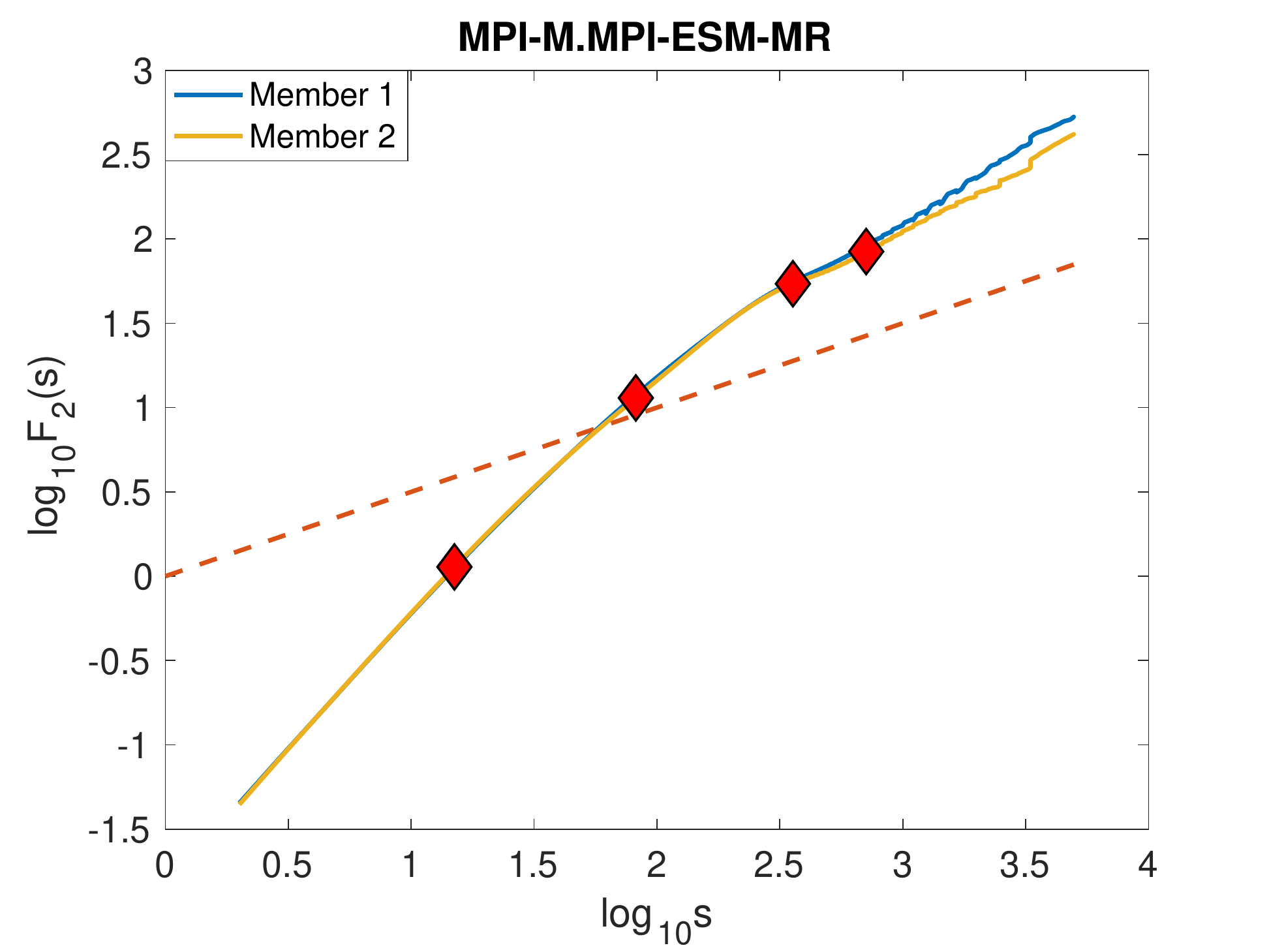}
   
   (c)\includegraphics[width = 0.45\textwidth]{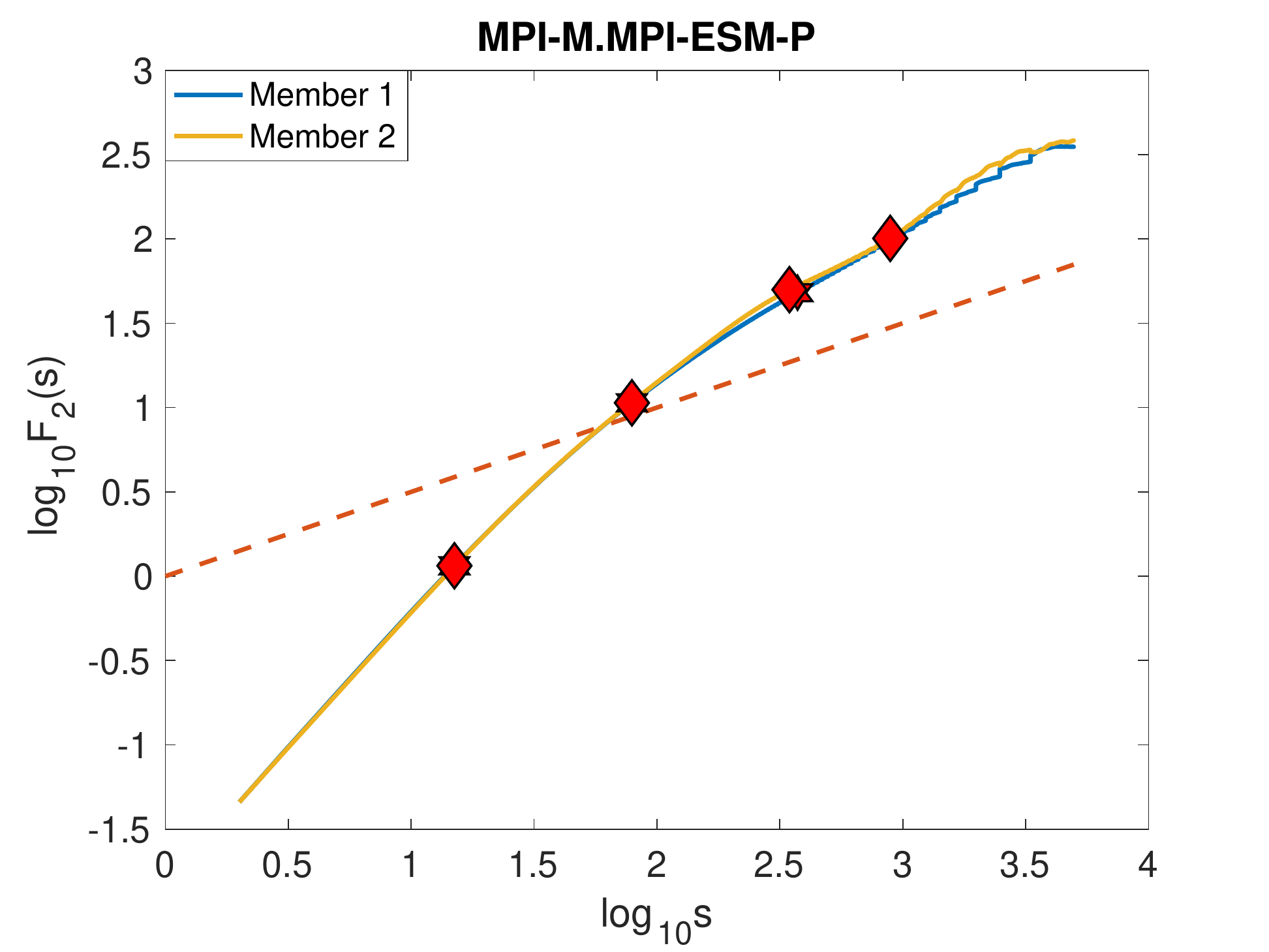}
   (d)\includegraphics[width = 0.45\textwidth]{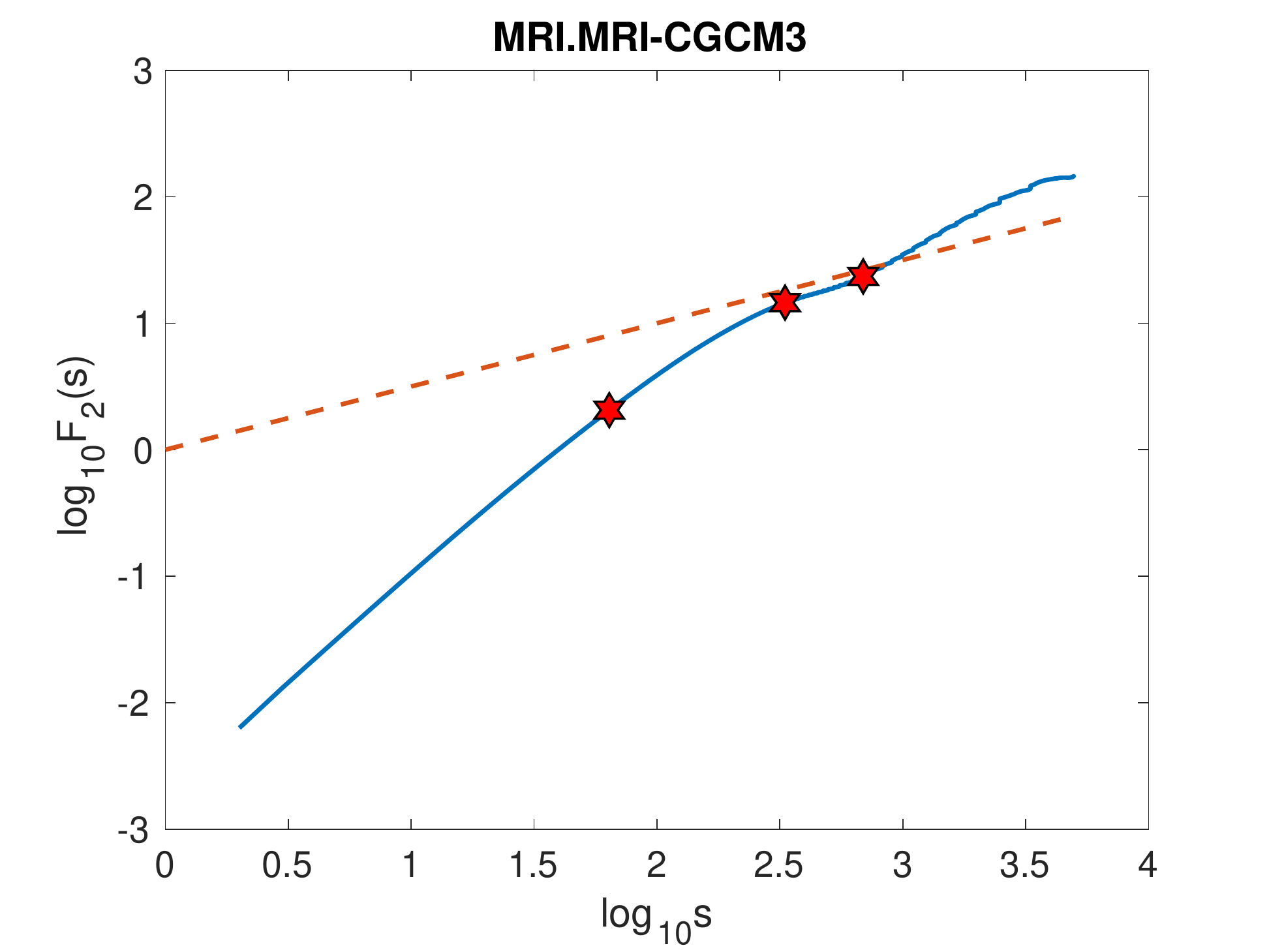}
   
   (e)\includegraphics[width = 0.45\textwidth]{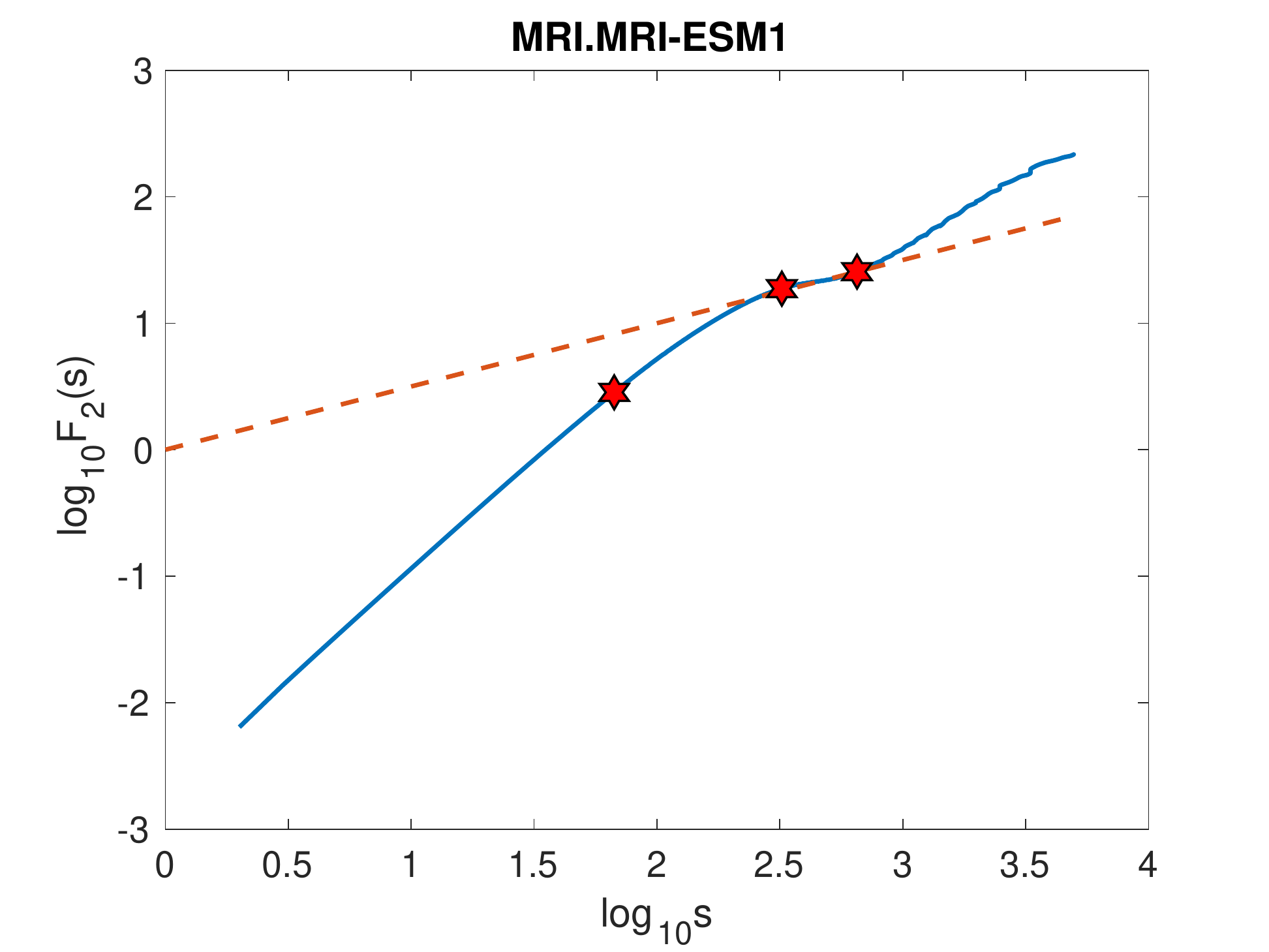}
   (f)\includegraphics[width = 0.45\textwidth]{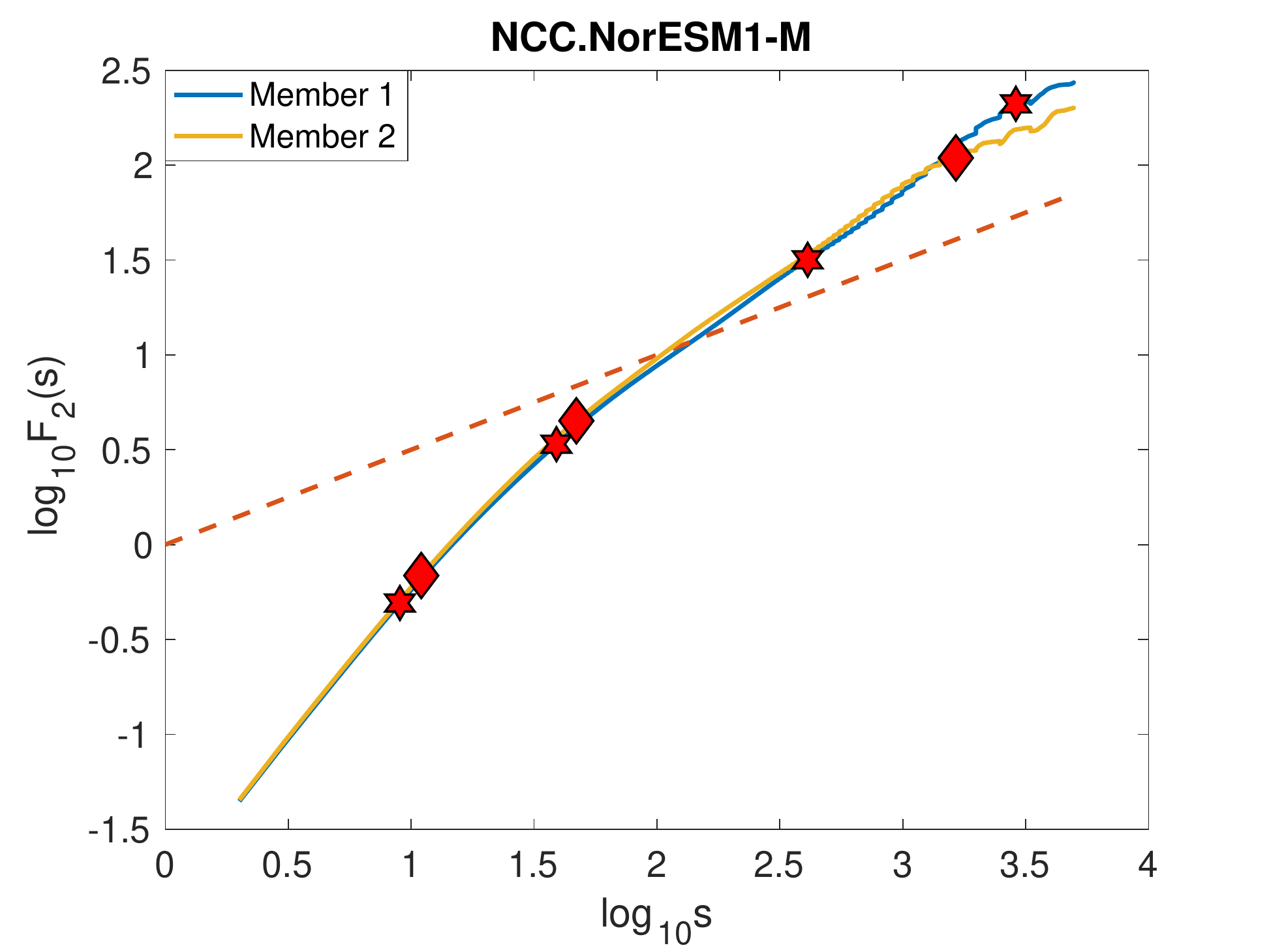}   
\caption{Fluctuation functions (blue) for CMIP5 AR5 models. The respective timescales present are (a) { Member 1: 15 days, 328 days, 760 days (stars); Member 2: 15 days, 332 days, 768 days (diamonds); (b) Member 1: 15 days, 82 days, 357 days, 708 days (stars); Member 2: 15 days, 82 days, 357 days, 710 days (diamonds); (c) Member 1: 15 days, 79 days, 373 days (stars); Member 2: 15 days, 79 days, 346 days, 889 days (diamonds); }(d) 64 days, 332 days, 690 days (stars); (e) 67 days, 322 days, 652 days (stars); (f) { Member 1: 9 days, 39 days, 410 days, 7.9 years (stars); Member 2: 11 days, 47 days, 4.5 years (diamonds)}. The dashed line (orange) denotes a slope of 1/2.}
\label{fig:GCM_3}
\end{figure*}

\begin{figure*}
\centering
   (a)\includegraphics[width = 0.45\textwidth]{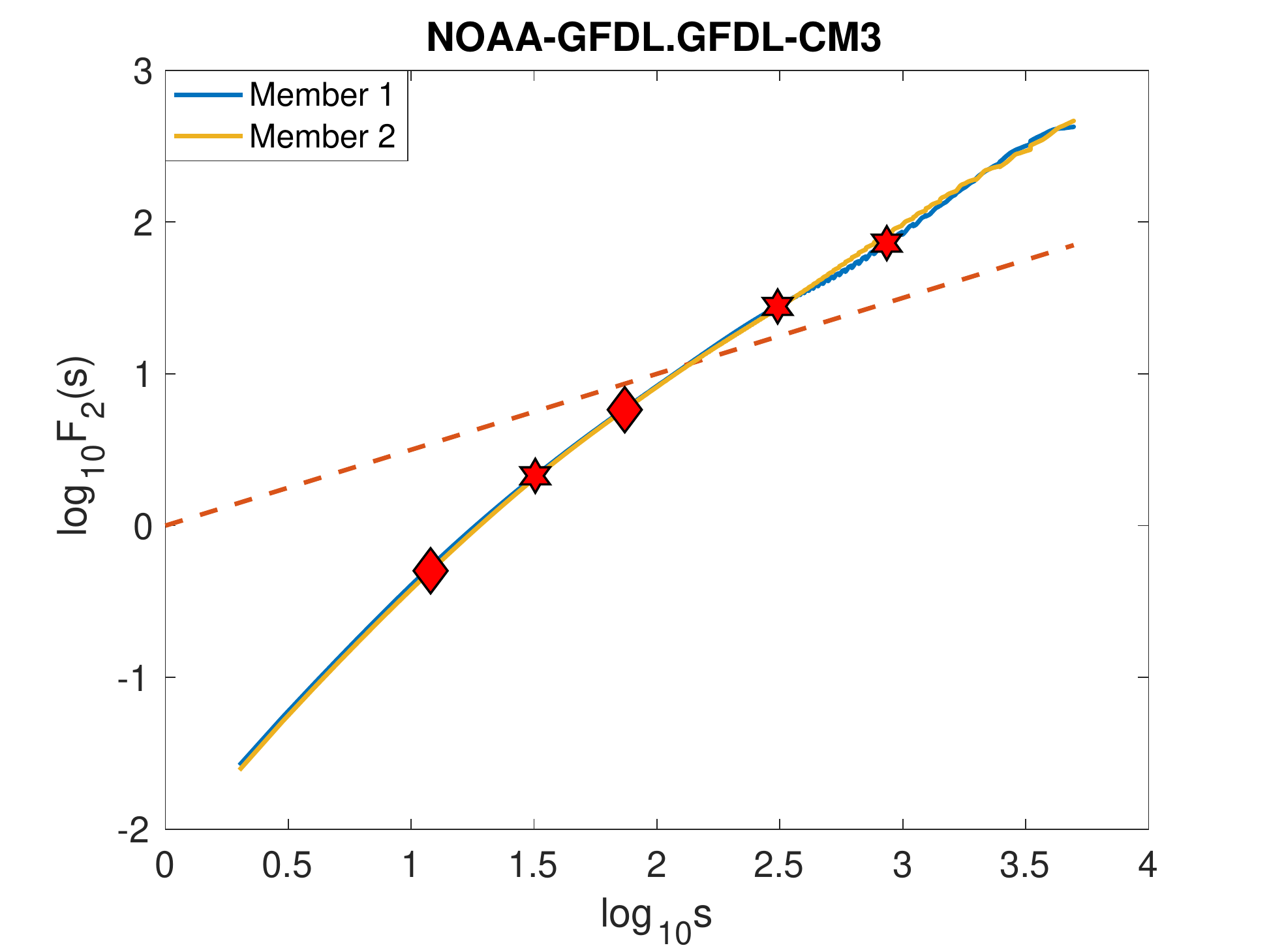}
   (b)\includegraphics[width = 0.45\textwidth]{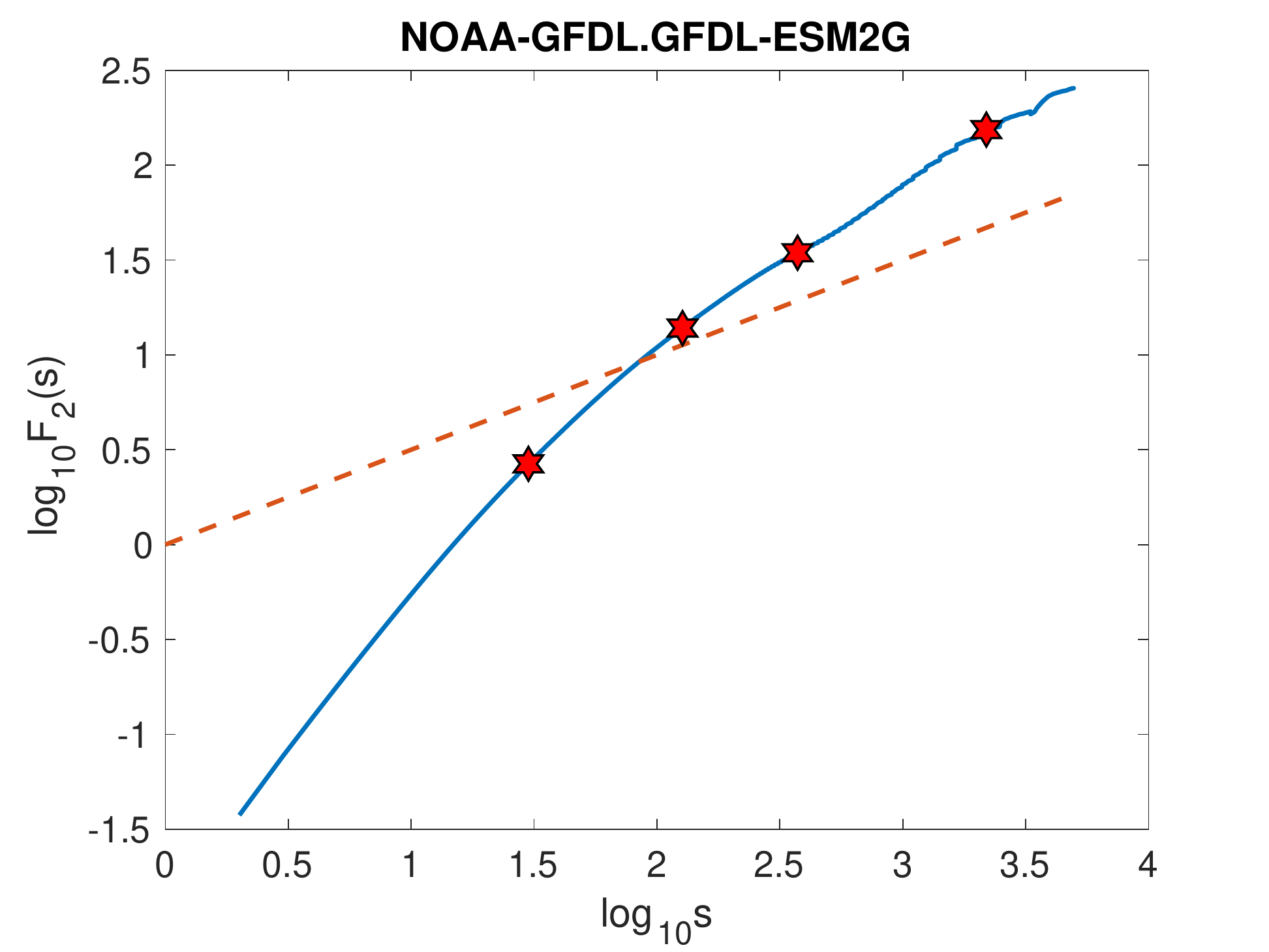}
   
   (c)\includegraphics[width = 0.45\textwidth]{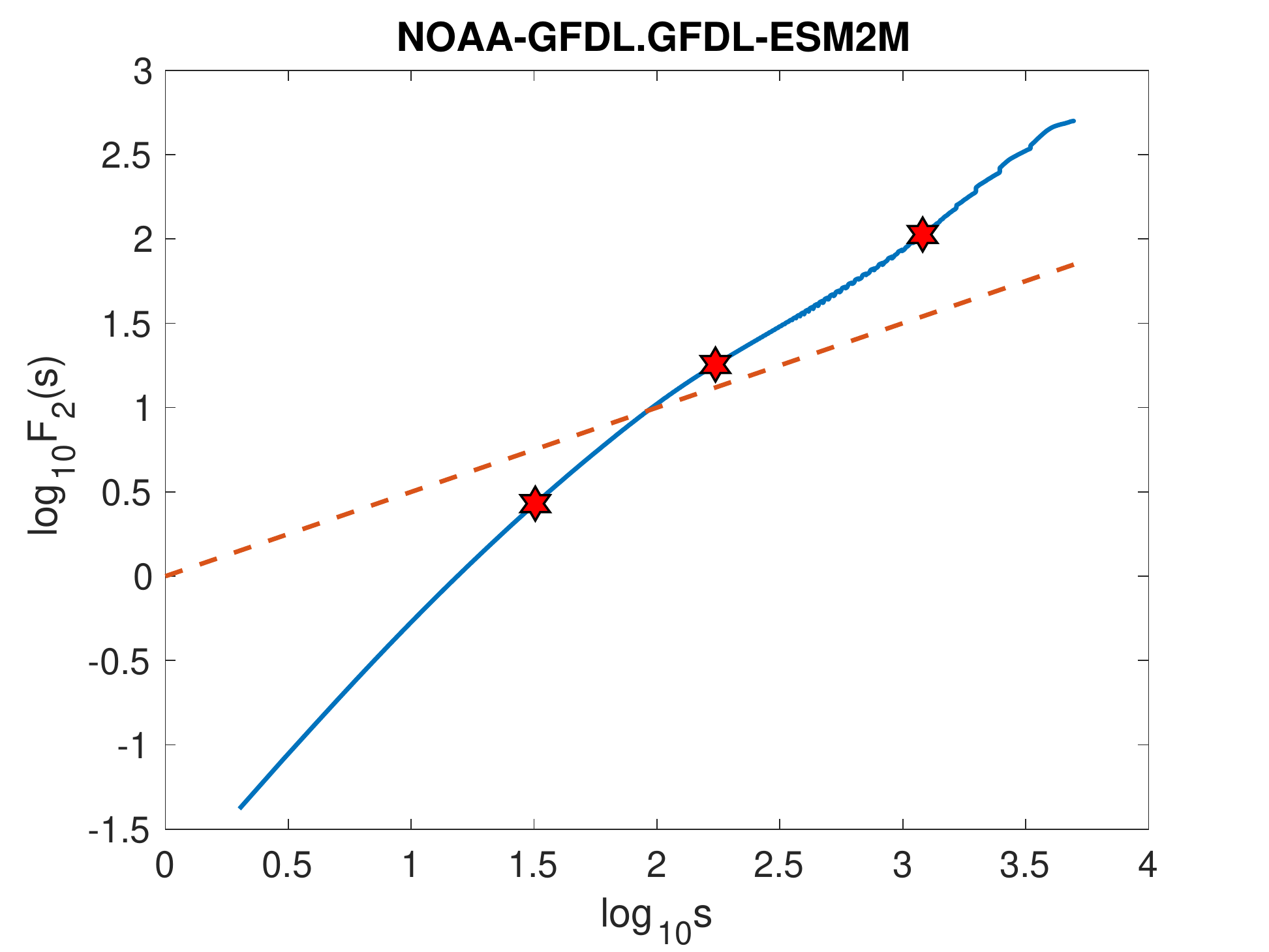}
\caption{Fluctuation functions (blue) for CMIP5 AR5 models. The respective timescales present are (a) { Member 1: 32 days, 310 days, 861 days (stars); Member 2: 12 days, 74 days (diamonds); }(b) 30 days, 127 days, 373 days, 6 years (stars); (c) 32 days, 173 days, 3.3 years (stars). The dashed line (orange) denotes a slope of 1/2.}
\label{fig:GCM_4}
\end{figure*}

Below we provide additional figures (\ref{fig:GCM_1} -- \ref{fig:GCM_4}) of the fluctuation functions, $F_q(s)$,  versus timescale, $s$, for the CMIP5 models labeled therein and described in Table \ref{Tb:models}.  Changes in the slope of $F_q(s)$ occur at various particular timescales denoted in the caption.  The main text describes the dynamics associated with these time scales.  

\begin{thebibliography}{32}%
\makeatletter
\providecommand \@ifxundefined [1]{%
 \@ifx{#1\undefined}
}%
\providecommand \@ifnum [1]{%
 \ifnum #1\expandafter \@firstoftwo
 \else \expandafter \@secondoftwo
 \fi
}%
\providecommand \@ifx [1]{%
 \ifx #1\expandafter \@firstoftwo
 \else \expandafter \@secondoftwo
 \fi
}%
\providecommand \natexlab [1]{#1}%
\providecommand \enquote  [1]{``#1''}%
\providecommand \bibnamefont  [1]{#1}%
\providecommand \bibfnamefont [1]{#1}%
\providecommand \citenamefont [1]{#1}%
\providecommand \href@noop [0]{\@secondoftwo}%
\providecommand \href [0]{\begingroup \@sanitize@url \@href}%
\providecommand \@href[1]{\@@startlink{#1}\@@href}%
\providecommand \@@href[1]{\endgroup#1\@@endlink}%
\providecommand \@sanitize@url [0]{\catcode `\\12\catcode `\$12\catcode
  `\&12\catcode `\#12\catcode `\^12\catcode `\_12\catcode `\%12\relax}%
\providecommand \@@startlink[1]{}%
\providecommand \@@endlink[0]{}%
\providecommand \url  [0]{\begingroup\@sanitize@url \@url }%
\providecommand \@url [1]{\endgroup\@href {#1}{\urlprefix }}%
\providecommand \urlprefix  [0]{URL }%
\providecommand \Eprint [0]{\href }%
\providecommand \doibase [0]{http://dx.doi.org/}%
\providecommand \selectlanguage [0]{\@gobble}%
\providecommand \bibinfo  [0]{\@secondoftwo}%
\providecommand \bibfield  [0]{\@secondoftwo}%
\providecommand \translation [1]{[#1]}%
\providecommand \BibitemOpen [0]{}%
\providecommand \bibitemStop [0]{}%
\providecommand \bibitemNoStop [0]{.\EOS\space}%
\providecommand \EOS [0]{\spacefactor3000\relax}%
\providecommand \BibitemShut  [1]{\csname bibitem#1\endcsname}%
\let\auto@bib@innerbib\@empty
\bibitem [{\citenamefont {Kwok}\ and\ \citenamefont
  {Untersteiner}(2011)}]{OneWatt}%
  \BibitemOpen
  \bibfield  {author} {\bibinfo {author} {\bibfnamefont {R.}~\bibnamefont
  {Kwok}}\ and\ \bibinfo {author} {\bibfnamefont {N.}~\bibnamefont
  {Untersteiner}},\ }\href@noop {} {\bibfield  {journal} {\bibinfo  {journal}
  {Phys. Today}\ }\textbf {\bibinfo {volume} {64}},\ \bibinfo {pages} {36}
  (\bibinfo {year} {2011})}\BibitemShut {NoStop}%
\bibitem [{\citenamefont {IPCC}(2013)}]{IPCC:2013}%
  \BibitemOpen
  \bibfield  {author} {\bibinfo {author} {\bibnamefont {IPCC}},\ }\enquote
  {\bibinfo {title} {Evaluation of climate models},}\ in\ \href@noop {} {\emph
  {\bibinfo {booktitle} {Climate Change 2013 --The Physical Science Basis:
  Working Group I Contribution to the Fifth Assessment Report of the
  Intergovernmental Panel on Climate Change}}}\ (\bibinfo  {publisher}
  {Cambridge University Press},\ \bibinfo {address} {Cambridge},\ \bibinfo
  {year} {2013})\ pp.\ \bibinfo {pages} {741--866}\BibitemShut {NoStop}%
\bibitem [{\citenamefont {Stroeve}\ \emph {et~al.}(2012)\citenamefont
  {Stroeve}, \citenamefont {Kattsov}, \citenamefont {Barrett}, \citenamefont
  {Serreze}, \citenamefont {Pavlova}, \citenamefont {Holland},\ and\
  \citenamefont {Meier}}]{Stroeve:2012aa}%
  \BibitemOpen
  \bibfield  {author} {\bibinfo {author} {\bibfnamefont {J.~C.}\ \bibnamefont
  {Stroeve}}, \bibinfo {author} {\bibfnamefont {V.}~\bibnamefont {Kattsov}},
  \bibinfo {author} {\bibfnamefont {A.}~\bibnamefont {Barrett}}, \bibinfo
  {author} {\bibfnamefont {M.}~\bibnamefont {Serreze}}, \bibinfo {author}
  {\bibfnamefont {T.}~\bibnamefont {Pavlova}}, \bibinfo {author} {\bibfnamefont
  {M.}~\bibnamefont {Holland}}, \ and\ \bibinfo {author} {\bibfnamefont
  {W.~N.}\ \bibnamefont {Meier}},\ }\href@noop {} {\bibfield  {journal}
  {\bibinfo  {journal} {Geophys. Res. Lett.}\ }\textbf {\bibinfo {volume}
  {39}},\ \bibinfo {pages} {L16502} (\bibinfo {year} {2012})}\BibitemShut
  {NoStop}%
\bibitem [{\citenamefont {Wang}\ and\ \citenamefont
  {Overland}(2012)}]{Wang:2012}%
  \BibitemOpen
  \bibfield  {author} {\bibinfo {author} {\bibfnamefont {M.}~\bibnamefont
  {Wang}}\ and\ \bibinfo {author} {\bibfnamefont {J.~E.}\ \bibnamefont
  {Overland}},\ }\href@noop {} {\bibfield  {journal} {\bibinfo  {journal}
  {Geophys. Res. Lett.}\ }\textbf {\bibinfo {volume} {39}},\ \bibinfo {pages}
  {L18501} (\bibinfo {year} {2012})}\BibitemShut {NoStop}%
\bibitem [{\citenamefont {Eisenman}\ \emph {et~al.}(2007)\citenamefont
  {Eisenman}, \citenamefont {Untersteiner},\ and\ \citenamefont
  {Wettlaufer}}]{EUW:2007}%
  \BibitemOpen
  \bibfield  {author} {\bibinfo {author} {\bibfnamefont {I.}~\bibnamefont
  {Eisenman}}, \bibinfo {author} {\bibfnamefont {N.}~\bibnamefont
  {Untersteiner}}, \ and\ \bibinfo {author} {\bibfnamefont {J.~S.}\
  \bibnamefont {Wettlaufer}},\ }\href@noop {} {\bibfield  {journal} {\bibinfo
  {journal} {Geophys. Res. Lett.}\ }\textbf {\bibinfo {volume} {34}},\ \bibinfo
  {pages} {L10501} (\bibinfo {year} {2007})}\BibitemShut {NoStop}%
\bibitem [{\citenamefont {DeWeaver}\ \emph {et~al.}(2008)\citenamefont
  {DeWeaver}, \citenamefont {Hunke},\ and\ \citenamefont
  {Holland}}]{DeWeaver:2008}%
  \BibitemOpen
  \bibfield  {author} {\bibinfo {author} {\bibfnamefont {E.~T.}\ \bibnamefont
  {DeWeaver}}, \bibinfo {author} {\bibfnamefont {E.~C.}\ \bibnamefont {Hunke}},
  \ and\ \bibinfo {author} {\bibfnamefont {M.~M.}\ \bibnamefont {Holland}},\
  }\href@noop {} {\bibfield  {journal} {\bibinfo  {journal} {Geophys. Res.
  Lett.}\ }\textbf {\bibinfo {volume} {35}},\ \bibinfo {pages} {L04501}
  (\bibinfo {year} {2008})}\BibitemShut {NoStop}%
\bibitem [{\citenamefont {Eisenman}\ \emph {et~al.}(2008)\citenamefont
  {Eisenman}, \citenamefont {Untersteiner},\ and\ \citenamefont
  {Wettlaufer}}]{EUW:2008}%
  \BibitemOpen
  \bibfield  {author} {\bibinfo {author} {\bibfnamefont {I.}~\bibnamefont
  {Eisenman}}, \bibinfo {author} {\bibfnamefont {N.}~\bibnamefont
  {Untersteiner}}, \ and\ \bibinfo {author} {\bibfnamefont {J.~S.}\
  \bibnamefont {Wettlaufer}},\ }\href@noop {} {\bibfield  {journal} {\bibinfo
  {journal} {Geophys. Res. Lett.}\ }\textbf {\bibinfo {volume} {35}},\ \bibinfo
  {pages} {L04502} (\bibinfo {year} {2008})}\BibitemShut {NoStop}%
\bibitem [{\citenamefont {Mauritsen}\ \emph {et~al.}(2012)\citenamefont
  {Mauritsen}, \citenamefont {Stevens}, \citenamefont {Roeckner}, \citenamefont
  {Crueger}, \citenamefont {Esch}, \citenamefont {Giorgetta}, \citenamefont
  {Haak}, \citenamefont {Jungclaus}, \citenamefont {Klocke}, \citenamefont
  {Matei}, \citenamefont {Mikolajewicz}, \citenamefont {Notz}, \citenamefont
  {Pincus}, \citenamefont {Schmidt},\ and\ \citenamefont
  {Tomassini}}]{Mauritsen:2012aa}%
  \BibitemOpen
  \bibfield  {author} {\bibinfo {author} {\bibfnamefont {T.}~\bibnamefont
  {Mauritsen}}, \bibinfo {author} {\bibfnamefont {B.}~\bibnamefont {Stevens}},
  \bibinfo {author} {\bibfnamefont {E.}~\bibnamefont {Roeckner}}, \bibinfo
  {author} {\bibfnamefont {T.}~\bibnamefont {Crueger}}, \bibinfo {author}
  {\bibfnamefont {M.}~\bibnamefont {Esch}}, \bibinfo {author} {\bibfnamefont
  {M.}~\bibnamefont {Giorgetta}}, \bibinfo {author} {\bibfnamefont
  {H.}~\bibnamefont {Haak}}, \bibinfo {author} {\bibfnamefont {J.}~\bibnamefont
  {Jungclaus}}, \bibinfo {author} {\bibfnamefont {D.}~\bibnamefont {Klocke}},
  \bibinfo {author} {\bibfnamefont {D.}~\bibnamefont {Matei}}, \bibinfo
  {author} {\bibfnamefont {U.}~\bibnamefont {Mikolajewicz}}, \bibinfo {author}
  {\bibfnamefont {D.}~\bibnamefont {Notz}}, \bibinfo {author} {\bibfnamefont
  {R.}~\bibnamefont {Pincus}}, \bibinfo {author} {\bibfnamefont
  {H.}~\bibnamefont {Schmidt}}, \ and\ \bibinfo {author} {\bibfnamefont
  {L.}~\bibnamefont {Tomassini}},\ }\href@noop {} {\bibfield  {journal}
  {\bibinfo  {journal} {J. Adv. Model. Earth Syst.}\ }\textbf {\bibinfo
  {volume} {4}},\ \bibinfo {pages} {M00A01} (\bibinfo {year}
  {2012})}\BibitemShut {NoStop}%
\bibitem [{\citenamefont {Hourdin}\ \emph {et~al.}(2013)\citenamefont
  {Hourdin}, \citenamefont {Grandpeix}, \citenamefont {Rio}, \citenamefont
  {Bony}, \citenamefont {Jam}, \citenamefont {Cheruy}, \citenamefont
  {Rochetin}, \citenamefont {Fairhead}, \citenamefont {Idelkadi}, \citenamefont
  {Musat}, \citenamefont {Dufresne}, \citenamefont {Lahellec}, \citenamefont
  {Lefebvre},\ and\ \citenamefont {Roehrig}}]{Hourdin:2013aa}%
  \BibitemOpen
  \bibfield  {author} {\bibinfo {author} {\bibfnamefont {F.}~\bibnamefont
  {Hourdin}}, \bibinfo {author} {\bibfnamefont {J.-Y.}\ \bibnamefont
  {Grandpeix}}, \bibinfo {author} {\bibfnamefont {C.}~\bibnamefont {Rio}},
  \bibinfo {author} {\bibfnamefont {S.}~\bibnamefont {Bony}}, \bibinfo {author}
  {\bibfnamefont {A.}~\bibnamefont {Jam}}, \bibinfo {author} {\bibfnamefont
  {F.}~\bibnamefont {Cheruy}}, \bibinfo {author} {\bibfnamefont
  {N.}~\bibnamefont {Rochetin}}, \bibinfo {author} {\bibfnamefont
  {L.}~\bibnamefont {Fairhead}}, \bibinfo {author} {\bibfnamefont
  {A.}~\bibnamefont {Idelkadi}}, \bibinfo {author} {\bibfnamefont
  {I.}~\bibnamefont {Musat}}, \bibinfo {author} {\bibfnamefont {J.-L.}\
  \bibnamefont {Dufresne}}, \bibinfo {author} {\bibfnamefont {A.}~\bibnamefont
  {Lahellec}}, \bibinfo {author} {\bibfnamefont {M.-P.}\ \bibnamefont
  {Lefebvre}}, \ and\ \bibinfo {author} {\bibfnamefont {R.}~\bibnamefont
  {Roehrig}},\ }\href@noop {} {\bibfield  {journal} {\bibinfo  {journal} {Clim.
  Dyn.}\ }\textbf {\bibinfo {volume} {40}},\ \bibinfo {pages} {2193} (\bibinfo
  {year} {2013})}\BibitemShut {NoStop}%
\bibitem [{\citenamefont {Agarwal}\ \emph {et~al.}(2012)\citenamefont
  {Agarwal}, \citenamefont {Moon},\ and\ \citenamefont
  {Wettlaufer}}]{Sahil:MF}%
  \BibitemOpen
  \bibfield  {author} {\bibinfo {author} {\bibfnamefont {S.}~\bibnamefont
  {Agarwal}}, \bibinfo {author} {\bibfnamefont {W.}~\bibnamefont {Moon}}, \
  and\ \bibinfo {author} {\bibfnamefont {J.~S.}\ \bibnamefont {Wettlaufer}},\
  }\href@noop {} {\bibfield  {journal} {\bibinfo  {journal} {Proc. Roy. Soc.
  Lond. A}\ }\textbf {\bibinfo {volume} {468}},\ \bibinfo {pages} {2416}
  (\bibinfo {year} {2012})}\BibitemShut {NoStop}%
\bibitem [{\citenamefont {Eisenman}(2010)}]{IanGeom}%
  \BibitemOpen
  \bibfield  {author} {\bibinfo {author} {\bibfnamefont {I.}~\bibnamefont
  {Eisenman}},\ }\href@noop {} {\bibfield  {journal} {\bibinfo  {journal}
  {Geophys. Res. Lett.}\ }\textbf {\bibinfo {volume} {37}},\ \bibinfo {pages}
  {L16501} (\bibinfo {year} {2010})}\BibitemShut {NoStop}%
\bibitem [{\citenamefont {Piwowar}\ \emph {et~al.}(1996)\citenamefont
  {Piwowar}, \citenamefont {Wessel},\ and\ \citenamefont
  {LeDrew}}]{Piwowar:1996}%
  \BibitemOpen
  \bibfield  {author} {\bibinfo {author} {\bibfnamefont {J.~M.}\ \bibnamefont
  {Piwowar}}, \bibinfo {author} {\bibfnamefont {G.~R.}\ \bibnamefont {Wessel}},
  \ and\ \bibinfo {author} {\bibfnamefont {E.~F.}\ \bibnamefont {LeDrew}},\
  }\href@noop {} {\emph {\bibinfo {title} {Geoscience and Remote Sensing
  Symposium, 1996. IGARSS'96.'Remote Sensing for a Sustainable Future.',
  International}}},\ Vol.~\bibinfo {volume} {1}\ (\bibinfo  {publisher}
  {IEEE},\ \bibinfo {year} {1996})\ pp.\ \bibinfo {pages}
  {645--647}\BibitemShut {NoStop}%
\bibitem [{\citenamefont {Blanchard-Wrigglesworth}\ \emph
  {et~al.}(2010)\citenamefont {Blanchard-Wrigglesworth}, \citenamefont
  {Armour}, \citenamefont {Bitz},\ and\ \citenamefont
  {DeWeaver}}]{Blanchard:2010aa}%
  \BibitemOpen
  \bibfield  {author} {\bibinfo {author} {\bibfnamefont {E.}~\bibnamefont
  {Blanchard-Wrigglesworth}}, \bibinfo {author} {\bibfnamefont {K.~C.}\
  \bibnamefont {Armour}}, \bibinfo {author} {\bibfnamefont {C.~M.}\
  \bibnamefont {Bitz}}, \ and\ \bibinfo {author} {\bibfnamefont
  {E.}~\bibnamefont {DeWeaver}},\ }\href@noop {} {\bibfield  {journal}
  {\bibinfo  {journal} {J. Clim.}\ }\textbf {\bibinfo {volume} {24}},\ \bibinfo
  {pages} {231} (\bibinfo {year} {2010})}\BibitemShut {NoStop}%
\bibitem [{\citenamefont {Armour}\ \emph {et~al.}(2011)\citenamefont {Armour},
  \citenamefont {Bitz}, \citenamefont {Thompson},\ and\ \citenamefont
  {Hunke}}]{Armour:2011}%
  \BibitemOpen
  \bibfield  {author} {\bibinfo {author} {\bibfnamefont {K.~C.}\ \bibnamefont
  {Armour}}, \bibinfo {author} {\bibfnamefont {C.~M.}\ \bibnamefont {Bitz}},
  \bibinfo {author} {\bibfnamefont {L.}~\bibnamefont {Thompson}}, \ and\
  \bibinfo {author} {\bibfnamefont {E.~C.}\ \bibnamefont {Hunke}},\ }\href@noop
  {} {\bibfield  {journal} {\bibinfo  {journal} {J. Clim.}\ }\textbf {\bibinfo
  {volume} {24}},\ \bibinfo {pages} {2378} (\bibinfo {year}
  {2011})}\BibitemShut {NoStop}%
\bibitem [{\citenamefont {Lindsay}\ \emph {et~al.}(2008)\citenamefont
  {Lindsay}, \citenamefont {Zhang}, \citenamefont {Schweiger},\ and\
  \citenamefont {Steele}}]{Lindsay:2008}%
  \BibitemOpen
  \bibfield  {author} {\bibinfo {author} {\bibfnamefont {R.~W.}\ \bibnamefont
  {Lindsay}}, \bibinfo {author} {\bibfnamefont {J.}~\bibnamefont {Zhang}},
  \bibinfo {author} {\bibfnamefont {A.~J.}\ \bibnamefont {Schweiger}}, \ and\
  \bibinfo {author} {\bibfnamefont {M.~A.}\ \bibnamefont {Steele}},\
  }\href@noop {} {\bibfield  {journal} {\bibinfo  {journal} {J. Geophys.
  Res.-Oceans}\ }\textbf {\bibinfo {volume} {113}},\ \bibinfo {pages} {C02023}
  (\bibinfo {year} {2008})}\BibitemShut {NoStop}%
\bibitem [{\citenamefont {Wang}\ \emph {et~al.}(2012)\citenamefont {Wang},
  \citenamefont {Chen},\ and\ \citenamefont {Kumar}}]{Wang:2012aa}%
  \BibitemOpen
  \bibfield  {author} {\bibinfo {author} {\bibfnamefont {W.}~\bibnamefont
  {Wang}}, \bibinfo {author} {\bibfnamefont {M.}~\bibnamefont {Chen}}, \ and\
  \bibinfo {author} {\bibfnamefont {A.}~\bibnamefont {Kumar}},\ }\href@noop {}
  {\bibfield  {journal} {\bibinfo  {journal} {Mon. Weather Rev.}\ }\textbf
  {\bibinfo {volume} {141}},\ \bibinfo {pages} {1375} (\bibinfo {year}
  {2012})}\BibitemShut {NoStop}%
\bibitem [{\citenamefont {Chevallier}\ \emph {et~al.}(2013)\citenamefont
  {Chevallier}, \citenamefont {Salas~y M{\'e}lia}, \citenamefont {Voldoire},
  \citenamefont {D{\'e}qu{\'e}},\ and\ \citenamefont
  {Garric}}]{Chevallier:2013aa}%
  \BibitemOpen
  \bibfield  {author} {\bibinfo {author} {\bibfnamefont {M.}~\bibnamefont
  {Chevallier}}, \bibinfo {author} {\bibfnamefont {D.}~\bibnamefont {Salas~y
  M{\'e}lia}}, \bibinfo {author} {\bibfnamefont {A.}~\bibnamefont {Voldoire}},
  \bibinfo {author} {\bibfnamefont {M.}~\bibnamefont {D{\'e}qu{\'e}}}, \ and\
  \bibinfo {author} {\bibfnamefont {G.}~\bibnamefont {Garric}},\ }\href@noop {}
  {\bibfield  {journal} {\bibinfo  {journal} {J. Clim.}\ }\textbf {\bibinfo
  {volume} {26}},\ \bibinfo {pages} {6092} (\bibinfo {year}
  {2013})}\BibitemShut {NoStop}%
\bibitem [{\citenamefont {Sigmond}\ \emph {et~al.}(2013)\citenamefont
  {Sigmond}, \citenamefont {Fyfe}, \citenamefont {Flato}, \citenamefont
  {Kharin},\ and\ \citenamefont {Merryfield}}]{Sigmond:2013aa}%
  \BibitemOpen
  \bibfield  {author} {\bibinfo {author} {\bibfnamefont {M.}~\bibnamefont
  {Sigmond}}, \bibinfo {author} {\bibfnamefont {J.~C.}\ \bibnamefont {Fyfe}},
  \bibinfo {author} {\bibfnamefont {G.~M.}\ \bibnamefont {Flato}}, \bibinfo
  {author} {\bibfnamefont {V.~V.}\ \bibnamefont {Kharin}}, \ and\ \bibinfo
  {author} {\bibfnamefont {W.~J.}\ \bibnamefont {Merryfield}},\ }\href@noop {}
  {\bibfield  {journal} {\bibinfo  {journal} {Geophys. Res. Lett.}\ }\textbf
  {\bibinfo {volume} {40}},\ \bibinfo {pages} {529} (\bibinfo {year}
  {2013})}\BibitemShut {NoStop}%
\bibitem [{\citenamefont {Tietsche}\ \emph {et~al.}(2014)\citenamefont
  {Tietsche}, \citenamefont {Day}, \citenamefont {Guemas}, \citenamefont
  {Hurlin}, \citenamefont {Keeley}, \citenamefont {Matei}, \citenamefont
  {Msadek}, \citenamefont {Collins},\ and\ \citenamefont
  {Hawkins}}]{Tietsche:2014aa}%
  \BibitemOpen
  \bibfield  {author} {\bibinfo {author} {\bibfnamefont {S.}~\bibnamefont
  {Tietsche}}, \bibinfo {author} {\bibfnamefont {J.~J.}\ \bibnamefont {Day}},
  \bibinfo {author} {\bibfnamefont {V.}~\bibnamefont {Guemas}}, \bibinfo
  {author} {\bibfnamefont {W.~J.}\ \bibnamefont {Hurlin}}, \bibinfo {author}
  {\bibfnamefont {S.~P.~E.}\ \bibnamefont {Keeley}}, \bibinfo {author}
  {\bibfnamefont {D.}~\bibnamefont {Matei}}, \bibinfo {author} {\bibfnamefont
  {R.}~\bibnamefont {Msadek}}, \bibinfo {author} {\bibfnamefont
  {M.}~\bibnamefont {Collins}}, \ and\ \bibinfo {author} {\bibfnamefont
  {E.}~\bibnamefont {Hawkins}},\ }\href@noop {} {\bibfield  {journal} {\bibinfo
   {journal} {Geophys. Res. Lett.}\ }\textbf {\bibinfo {volume} {41}},\
  \bibinfo {pages} {1035} (\bibinfo {year} {2014})}\BibitemShut {NoStop}%
\bibitem [{\citenamefont {Li}\ \emph {et~al.}(2017)\citenamefont {Li},
  \citenamefont {Zhang},\ and\ \citenamefont {Knutson}}]{Li:2017aa}%
  \BibitemOpen
  \bibfield  {author} {\bibinfo {author} {\bibfnamefont {D.}~\bibnamefont
  {Li}}, \bibinfo {author} {\bibfnamefont {R.}~\bibnamefont {Zhang}}, \ and\
  \bibinfo {author} {\bibfnamefont {T.~R.}\ \bibnamefont {Knutson}},\
  }\href@noop {} {\bibfield  {journal} {\bibinfo  {journal} {Nat. Commun.}\
  }\textbf {\bibinfo {volume} {8}} (\bibinfo {year} {2017})}\BibitemShut
  {NoStop}%
\bibitem [{\citenamefont {Kantelhardt}\ \emph {et~al.}(2002)\citenamefont
  {Kantelhardt}, \citenamefont {Zschiegner}, \citenamefont {Koscielny-Bunde},
  \citenamefont {Havlin}, \citenamefont {Bunde},\ and\ \citenamefont
  {Stanley}}]{Kantelhardt:2002}%
  \BibitemOpen
  \bibfield  {author} {\bibinfo {author} {\bibfnamefont {J.~W.}\ \bibnamefont
  {Kantelhardt}}, \bibinfo {author} {\bibfnamefont {S.~A.}\ \bibnamefont
  {Zschiegner}}, \bibinfo {author} {\bibfnamefont {E.}~\bibnamefont
  {Koscielny-Bunde}}, \bibinfo {author} {\bibfnamefont {S.}~\bibnamefont
  {Havlin}}, \bibinfo {author} {\bibfnamefont {A.}~\bibnamefont {Bunde}}, \
  and\ \bibinfo {author} {\bibfnamefont {H.~E.}\ \bibnamefont {Stanley}},\
  }\href@noop {} {\bibfield  {journal} {\bibinfo  {journal} {Physica A}\
  }\textbf {\bibinfo {volume} {316}},\ \bibinfo {pages} {87} (\bibinfo {year}
  {2002})}\BibitemShut {NoStop}%
\bibitem [{\citenamefont {Hurst}(1951)}]{Hurst:1951aa}%
  \BibitemOpen
  \bibfield  {author} {\bibinfo {author} {\bibfnamefont {H.~E.}\ \bibnamefont
  {Hurst}},\ }\href@noop {} {\bibfield  {journal} {\bibinfo  {journal} {Trans.
  Am. Soc. Civ. Eng.}\ }\textbf {\bibinfo {volume} {116}},\ \bibinfo {pages}
  {770} (\bibinfo {year} {1951})}\BibitemShut {NoStop}%
\bibitem [{\citenamefont {Zhou}\ and\ \citenamefont {Leung}(2010)}]{Zhou:2010}%
  \BibitemOpen
  \bibfield  {author} {\bibinfo {author} {\bibfnamefont {Y.}~\bibnamefont
  {Zhou}}\ and\ \bibinfo {author} {\bibfnamefont {Y.}~\bibnamefont {Leung}},\
  }\href@noop {} {\bibfield  {journal} {\bibinfo  {journal} {J. Stat. Mech.:
  Theor. Exp.}\ ,\ \bibinfo {pages} {P06021}} (\bibinfo {year}
  {2010})}\BibitemShut {NoStop}%
\bibitem [{\citenamefont {Agarwal}\ and\ \citenamefont
  {Wettlaufer}(2017{\natexlab{a}})}]{Sahil:SIV}%
  \BibitemOpen
  \bibfield  {author} {\bibinfo {author} {\bibfnamefont {S.}~\bibnamefont
  {Agarwal}}\ and\ \bibinfo {author} {\bibfnamefont {J.~S.}\ \bibnamefont
  {Wettlaufer}},\ }\href@noop {} {\bibfield  {journal} {\bibinfo  {journal} {J.
  Clim.}\ }\textbf {\bibinfo {volume} {30}},\ \bibinfo {pages} {4873} (\bibinfo
  {year} {2017}{\natexlab{a}})}\BibitemShut {NoStop}%
\bibitem [{\citenamefont {Agarwal}\ \emph {et~al.}(2017)\citenamefont
  {Agarwal}, \citenamefont {Sordo},\ and\ \citenamefont
  {Wettlaufer}}]{Sahil:EXO}%
  \BibitemOpen
  \bibfield  {author} {\bibinfo {author} {\bibfnamefont {S.}~\bibnamefont
  {Agarwal}}, \bibinfo {author} {\bibfnamefont {F.~D.}\ \bibnamefont {Sordo}},
  \ and\ \bibinfo {author} {\bibfnamefont {J.~S.}\ \bibnamefont {Wettlaufer}},\
  }\href@noop {} {\bibfield  {journal} {\bibinfo  {journal} {Astron. J.}\
  }\textbf {\bibinfo {volume} {153}},\ \bibinfo {pages} {12} (\bibinfo {year}
  {2017})}\BibitemShut {NoStop}%
\bibitem [{\citenamefont {Agarwal}\ and\ \citenamefont
  {Wettlaufer}(2017{\natexlab{b}})}]{Sahil:ExoAtmos}%
  \BibitemOpen
  \bibfield  {author} {\bibinfo {author} {\bibfnamefont {S.}~\bibnamefont
  {Agarwal}}\ and\ \bibinfo {author} {\bibfnamefont {J.~S.}\ \bibnamefont
  {Wettlaufer}},\ }\href@noop {} {\bibfield  {journal} {\bibinfo  {journal}
  {submitted (arXiv:1710.09870)}\ } (\bibinfo {year}
  {2017}{\natexlab{b}})}\BibitemShut {NoStop}%
\bibitem [{\citenamefont {Rangarajan}\ and\ \citenamefont {Ding}(2000)}]{Ding}%
  \BibitemOpen
  \bibfield  {author} {\bibinfo {author} {\bibfnamefont {G.}~\bibnamefont
  {Rangarajan}}\ and\ \bibinfo {author} {\bibfnamefont {M.}~\bibnamefont
  {Ding}},\ }\href@noop {} {\bibfield  {journal} {\bibinfo  {journal} {Phys.
  Rev. E}\ }\textbf {\bibinfo {volume} {61}},\ \bibinfo {pages} {4991}
  (\bibinfo {year} {2000})}\BibitemShut {NoStop}%
\bibitem [{\citenamefont {Zhang}\ and\ \citenamefont
  {Rothrock}(2003)}]{Zhang:2003aa}%
  \BibitemOpen
  \bibfield  {author} {\bibinfo {author} {\bibfnamefont {J.}~\bibnamefont
  {Zhang}}\ and\ \bibinfo {author} {\bibfnamefont {D.~A.}\ \bibnamefont
  {Rothrock}},\ }\href@noop {} {\bibfield  {journal} {\bibinfo  {journal} {Mon.
  Weather Rev.}\ }\textbf {\bibinfo {volume} {131}},\ \bibinfo {pages} {845}
  (\bibinfo {year} {2003})}\BibitemShut {NoStop}%
\bibitem [{\citenamefont {Shu}\ \emph {et~al.}(2015)\citenamefont {Shu},
  \citenamefont {Song},\ and\ \citenamefont {Qiao}}]{Shu:2015aa}%
  \BibitemOpen
  \bibfield  {author} {\bibinfo {author} {\bibfnamefont {Q.}~\bibnamefont
  {Shu}}, \bibinfo {author} {\bibfnamefont {Z.}~\bibnamefont {Song}}, \ and\
  \bibinfo {author} {\bibfnamefont {F.}~\bibnamefont {Qiao}},\ }\href@noop {}
  {\bibfield  {journal} {\bibinfo  {journal} {The Cryosphere}\ }\textbf
  {\bibinfo {volume} {9}},\ \bibinfo {pages} {399} (\bibinfo {year}
  {2015})}\BibitemShut {NoStop}%
\bibitem [{\citenamefont {Colony}\ and\ \citenamefont
  {Thorndike}(1980)}]{Colony:1980}%
  \BibitemOpen
  \bibfield  {author} {\bibinfo {author} {\bibfnamefont {R.}~\bibnamefont
  {Colony}}\ and\ \bibinfo {author} {\bibfnamefont {A.~S.}\ \bibnamefont
  {Thorndike}},\ }\href@noop {} {\bibfield  {journal} {\bibinfo  {journal} {J.
  Phys. Ocean.}\ }\textbf {\bibinfo {volume} {10}},\ \bibinfo {pages} {1281}
  (\bibinfo {year} {1980})}\BibitemShut {NoStop}%
\bibitem [{\citenamefont {Massonnet}\ \emph {et~al.}(2012)\citenamefont
  {Massonnet}, \citenamefont {Fichefet}, \citenamefont {Goosse}, \citenamefont
  {Bitz}, \citenamefont {Philippon-Berthier}, \citenamefont {Holland},\ and\
  \citenamefont {Barriat}}]{Massonnet:2012aa}%
  \BibitemOpen
  \bibfield  {author} {\bibinfo {author} {\bibfnamefont {F.}~\bibnamefont
  {Massonnet}}, \bibinfo {author} {\bibfnamefont {T.}~\bibnamefont {Fichefet}},
  \bibinfo {author} {\bibfnamefont {H.}~\bibnamefont {Goosse}}, \bibinfo
  {author} {\bibfnamefont {C.~M.}\ \bibnamefont {Bitz}}, \bibinfo {author}
  {\bibfnamefont {G.}~\bibnamefont {Philippon-Berthier}}, \bibinfo {author}
  {\bibfnamefont {M.~M.}\ \bibnamefont {Holland}}, \ and\ \bibinfo {author}
  {\bibfnamefont {P.~Y.}\ \bibnamefont {Barriat}},\ }\href@noop {} {\bibfield
  {journal} {\bibinfo  {journal} {The Cryosphere}\ }\textbf {\bibinfo {volume}
  {6}},\ \bibinfo {pages} {1383} (\bibinfo {year} {2012})}\BibitemShut
  {NoStop}%
\bibitem [{\citenamefont {Swart}\ \emph {et~al.}(2015)\citenamefont {Swart},
  \citenamefont {Fyfe}, \citenamefont {Hawkins}, \citenamefont {Kay},\ and\
  \citenamefont {Jahn}}]{Swart:2015aa}%
  \BibitemOpen
  \bibfield  {author} {\bibinfo {author} {\bibfnamefont {N.~C.}\ \bibnamefont
  {Swart}}, \bibinfo {author} {\bibfnamefont {J.~C.}\ \bibnamefont {Fyfe}},
  \bibinfo {author} {\bibfnamefont {E.}~\bibnamefont {Hawkins}}, \bibinfo
  {author} {\bibfnamefont {J.~E.}\ \bibnamefont {Kay}}, \ and\ \bibinfo
  {author} {\bibfnamefont {A.}~\bibnamefont {Jahn}},\ }\href@noop {} {\bibfield
   {journal} {\bibinfo  {journal} {Nature Clim. Change}\ }\textbf {\bibinfo
  {volume} {5}},\ \bibinfo {pages} {86} (\bibinfo {year} {2015})}\BibitemShut
  {NoStop}%
\end{thebibliography}
\end{document}